\documentclass[11pt,letterpaper]{JHEP3}
\usepackage{amsmath,epsfig,array,undertilde}
\usepackage{graphicx}
\usepackage{amssymb}
\usepackage{amsfonts,amstext,latexsym,xspace}
\usepackage{amsfonts}

\newcommand{\eq}{\begin{equation}}
\newcommand{\en}{\end{equation}}
\newcommand{\eqn}{\begin{eqnarray}}
\newcommand{\enn}{\end{eqnarray}}

\newcommand{\beq}{\begin{equation}}
\newcommand{\eeq}{\end{equation}}

\newcommand{\gd}[1][{}]{\delta_{#1}{}}

\newcommand{\cX}{\mathcal{X}}
\newcommand{\cY}{\mathcal{Y}}

\newcommand{\cN}{\mathcal{N}}

\title{Quasiconformal Realizations of $E_{6(6)}$, $E_{7(7)}$, $E_{8(8)}$ and $SO(n+3,m+3)$, 
$N \geqslant 4 $ Supergravity and Spherical Vectors}
\author{
Murat G\"{u}naydin$^{\dagger}$\footnote{murat@phys.psu.edu}
\\
$^{\dagger}$ \emph{Institute for Gravitation and the Cosmos \\ Physics Department \\
Penn  State University\\
University Park, PA 16802, USA} \\
}
\author{Oleksandr Pavlyk$^{\ddagger}$\footnote{pavlyk@wolfram.com}
\\
$^{\ddagger}$\emph{Wolfram Research Inc. \\100 Trade Center Dr. \\Champaign, IL 61820, USA} }

\abstract{ 
After reviewing the underlying algebraic structures we give a unified
realization of split exceptional groups $F_{4(4)}$,
$E_{6(6)}$, $E_{7(7)}$, $E_{8(8)}$ and of $SO(n+3,m+3)$ as quasiconformal groups that is covariant
with respect to their (Lorentz)  subgroups $SL(3,\mathbb{R})$, $SL(3,\mathbb{R})
\times SL(3,\mathbb{R})$, $SL(6,\mathbb{R})$, $E_{6(6)}$ and
$SO(n,m)\times SO(1,1)$, respectively. We determine the spherical
vectors of quasiconformal realizations of all these groups twisted by
a unitary character $\nu$. We also give their quadratic Casimir
operators and determine their values in terms of $\nu$ and the
dimension $n_V$ of the underlying Jordan algebras. For $\nu= -(n_V+2)
+i \rho$ the quasiconformal action induces unitary representations on
the space of square integrable functions in $(2n_V+3)$ variables, that
belong to the principle series. For special discrete values of $\nu$ the
quasiconformal action leads to unitary representations belonging to
the discrete series and their continuations. The 
manifolds that correspond to "quasiconformal compactifications" of the
respective $(2n_V+3)$ dimensional spaces are also given.  We discuss the relevance of
our results to $N=8$ supergravity and to $N=4$ Maxwell-Einstein
supergravity theories and , in particular, to the proposal that three
and four dimensional U-duality groups act as spectrum generating
quasiconformal and conformal groups of the corresponding four and five
dimensional supergravity theories, respectively. }
\keywords{Supergravity, Duality, Black Holes}
\preprint{}
\begin{document}
\renewcommand{\theequation}{\arabic{section}.{\arabic{subsection}}.\arabic{equation}}
\section{Introduction}
\setcounter{equation}{0} Earliest studies of unitary representations
of four dimensional U-duality groups of extended supergravity theories
were given in
\cite{Gunaydin:1981dc,Gunaydin:1981yq,Gunaydin:1981zm}. These
representations were constructed using oscillators that transform in
the same representations of the U-duality groups as the vector field
strengths plus their magnetic duals. These works were motivated
by the idea that in a quantum theory global symmetries
must be realized unitarily on the spectrum and the composite scenarios that attempted to connect maximal $N=8$
supergravity with observation
\cite{Ellis:1980tf,Ellis:1980cf,Gunaydin:1982gw}\footnote{
For further references on the subject, see \cite{Gunaydin:1982gw}}. 
In a composite scenario proposed in \cite{Ellis:1980tf} it was
conjectured that $SU(8)$ local symmetry of $N=8$ supergravity becomes
dynamical and acts as a family unifying grand unified theory (GUT)
which contains $SU(5)$ GUT as well as a family group $SU(3)$.  A
similar scenario leads to $E_6$ GUT with a family group $U(1)$ in the
exceptional supergravity theory \cite{Gunaydin:1983rk} whose U-duality
group is $E_{7(-25)}$ in $d=4$.  However with the discovery of possible counter
terms at higher loops it was argued that divergences would eventually
spoil the finiteness properties of $N=8$ supergravity.  After the work
of Green and Schwarz on anomaly cancellation in superstring theory
\cite{Green:1984sg} attempts at  composite scenarios in supergravity were
abandoned. However, recent discovery of unexpected cancellations of
divergences in supergravity theories \cite{Bern:2008pv,BjerrumBohr:2008dp,ArkaniHamed:2008gz,Chalmers:2000ks,Green:2006gt,Green:2006yu,Green:2007zzb,Kallosh:2008ru,Kallosh:2009db}
has brought back the question of finiteness of $N=8$ supergravity as
well as of exceptional supergravity.

Over the last decade or so, there has been a great deal more work done
on unitary representations of U-duality groups of extended
supergravity theories. The renewed interest in unitary realizations of
U-duality groups was due partly to the proposals that certain
extensions of U-duality groups may act as spectrum generating symmetry
groups of supergravity theories. Based on geometric considerations
involving orbits of extremal black hole solutions in $N=8$
supergravity and $N=2$ Maxwell-Einstein supergravity theories with
symmetric scalar manifolds, it was suggested that four dimensional
U-duality groups act as spectrum generating conformal symmetry groups
of the corresponding five dimensional supergravity theories
\cite{Ferrara:1997uz,Gunaydin:2000xr,Gunaydin:2004ku,Gunaydin:2003qm,Gunaydin:2005gd}. 
This proposal was then extended to the proposal that three dimensional
U-duality groups act as spectrum generating quasiconformal groups of
the corresponding four dimensional supergravity theories
\cite{Gunaydin:2000xr,Gunaydin:2004ku,Gunaydin:2003qm,Gunaydin:2005gd}. 
Quasiconformal realization of U-duality group $E_{8(8)}$ of three
dimensional maximal supergravity given in \cite{Gunaydin:2000xr} is
the first known geometric realization of $E_{8(8)}$ and its
quantization leads to the minimal unitary representation of $E_{8(8)}$
\cite{Gunaydin:2001bt}.  Remarkably, quasiconformal realizations exist
for different real forms of all noncompact groups and their
quantizations yield directly the minimal unitary representations of
the respective groups
\cite{Gunaydin:2001bt,Gunaydin:2004md,Gunaydin:2005zz,Gunaydin:2006vz}. 
Furthermore, the quasiconformal method gives a unified approach to the
minimal unitary representations of all noncompact groups and extends
also to supergroups \cite{Gunaydin:2006vz}. For symplectic groups
$Sp(2m,\mathbb{R})$ these minimal unitary representations are simply
the singleton representations.

Many results have been obtained over recent years that support the
proposals that four and three dimensional U-duality groups act as
spectrum generating conformal and quasiconformal groups of five and
four dimensional supergravity theories with symmetric scalar
manifolds, respectively. The work relating black hole solutions in
four and five dimensions (4d/5d lift)
\cite{Gaiotto:2005xt,Gaiotto:2005gf,Elvang:2005sa,Pioline:2005vi} is
consistent with the proposal that four dimensional U-duality groups
act as spectrum generating conformal symmetry groups of five
dimensional supergravity theories from which they
descend. Furthermore, the work of
\cite{Bouchareb:2007ax,Gal'tsov:2008nz} on using solution generating
techniques to relate the known black hole solutions of five
dimensional ungauged supergravity theories to each other and generate
new solutions using symmetry groups of the corresponding three
dimensional supergravity theories and related work on gauged
supergravity theories \cite{Berkooz:2008rj} are in accord with these
proposals.

A concrete framework for implementing the proposal that three
dimensional U-duality groups act as spectrum generating quasiconformal
groups was formulated  in
\cite{Gunaydin:2005mx,Gunaydin:2007bg,Gunaydin:2007qq} for spherically
symmetric stationary BPS black holes. This  framework is based on the fact that the
attractor equations \cite{Ferrara:1995ih,Ferrara:1996um} of spherically symmetric stationary black holes of
four dimensional supergravity theories are equivalent to the equations describing the geodesic
motion of a fiducial particle on the moduli space $\mathcal{M}^*_3$ of
three dimensional supergravity theories obtained by reduction on a
time-like circle\footnote{ This was first observed in
\cite{Breitenlohner:1987dg} and used in
\cite{Cvetic:1995kv,Cvetic:1995uj} to construct static and rotating
black holes in heterotic string theory.}. A related analysis on
non-BPS extremal black holes in theories with symmetric target
manifolds was carried out in \cite{Gaiotto:2007ag,Bergshoeff:2008be}.
For $N=2$ MESGTs
defined by Euclidean Jordan algebras of degree three the manifolds
$\mathcal{M}^*_3$ are para-quaternionic symmetric spaces
\begin{equation*}
 \mathcal{M}^*_3 =\frac{QConf(J)} { Conf(J)\times SU(1,1) }
\end{equation*}
where $QConf{J}$ and $Conf{J}$ are the quasiconformal and conformal
groups of the Jordan algebra $J$, respectively.  When one quantizes
the fiducial particle's motion one is led to quantum mechanical wave
functions that provide the basis of a unitary representation of
$QConf(J)$.  BPS black holes correspond to a special class of
geodesics
and the twistor space $\mathcal{Z}_3$ of $\mathcal{M}^*_3 $ can be
identified with the BPS phase space. Then the spherically symmetric
stationary BPS black hole solutions of $N=2$ MESGT's are described by
holomorphic curves in $\mathcal{Z}_3$
\cite{Gunaydin:2005mx,Gunaydin:2007bg,Gunaydin:2007qq,Neitzke:2007ke}.
One finds that the action of three dimensional U-duality group $QConf(J)$  on the
natural complex coordinates of the twistor space is
precisely of the quasiconformal form \cite{Gunaydin:2007qq}.
Therefore the unitary representations of $QConf(J)$ relevant for BPS
black holes of $N=2$ MESGT's  are those induced by holomorphic quasiconformal actions of
$QConf(J)$ on the corresponding twistor spaces $\mathcal{Z}_3$, which
belong in general to quaternionic discrete series representations of $QConf(J)$ \cite{Gunaydin:2007qq}.

Another result in support of the proposal that three dimensional
U-duality groups act as spectrum generating groups of the
corresponding four dimensional theories comes from the connection
established in \cite{Gunaydin:2007vc} between the harmonic superspace
(HSS) formulation of $N=2$, $d=4$ supersymmetric quaternionic K\"ahler
sigma models that couple to $N=2$ supergravity and the minimal unitary
representations of their isometry groups. One finds that for $N=2$
sigma models with quaternionic symmetric target spaces of the form\footnote{ $\widetilde{Conf}(J)$ is the compact form of $Conf(J)$.}
\[ \frac{QConf(J)}{\widetilde{Conf}(J)\times SU(2)}\] there exists a one-to-one
mapping between the quartic Killing potentials that generate the
isometry group $QConf(J)$ under Poisson brackets in the HSS
formulation, and the generators of the minimal unitary representation
of $QConf(J)$ obtained by quantization of its quasiconformal
realization. Therefore  the ``fundamental spectrum'' of the
quantum theory must fit into the minimal unitary representation of
$QConf(J)$ and the full spectrum is obtained by tensoring of the
minimal unitary representation.

In \cite{Gunaydin:2007qq} unitary representations of two quaternionic
groups of rank two, namely $SU(2,1)$ and $G_{2(2)}$, induced by their
geometric quasiconformal actions were studied in great detail. They
are the isometry groups of four and five dimensional simple $N=2$
supergravity theories dimensionally reduced on tori to three
dimensions, respectively.  Unitary representations induced by the
geometric quasiconformal action include the quaternionic discrete
series representations that were studied in mathematics literature
using other methods \cite{MR1421947}.  In the study of unitary
representations of $SU(2,1)$ and $G_{2(2)}$, in particular of
quaternionic discrete series, studied in \cite{Gunaydin:2007qq}
spherical vectors of maximal compact subgroups under their
quasiconformal actions play an essential role. In a recent paper
\cite{Gunaydin:2009dq} we gave a unified quasiconformal realization of
three dimensional U-duality groups $QConf(J)$ of all $N=2$ MESGTs with
symmetric scalar manifolds defined by Euclidean Jordan algebras of
degree three in a basis covariant with respect to their 5 dimensional
U-duality groups. These three dimensional U-duality groups are
$F_{4(4)}$, $E_{6(2)}$, $E_{7(-5)}$, $E_{8(-24)}$ and $SO(n_V+2,4)$ and
their five dimensional U-duality groups are $SL(3,\mathbb{R})$,
$SL(3,\mathbb{C})$, $SU^*(6)$, $E_{6(-26)}$ and $SO(1,1)\times
SO(n_V-1,1)$, respectively\footnote{Of course, the rank two
quaternionic quasiconformal groups can be obtained as a trivial limit
of the general unified formulation.}. We gave their quadratic Casimir
operators and determined their values and most importantly presented
the spherical vectors of all these quasiconformal groups in a unified
manner, which are essential for the construction of the quaternionic
discrete series representations.

In this paper we extend the results of \cite{Gunaydin:2009dq} to the
split exceptional groups $E_{6(6)}$, $E_{7(7)}$, $E_{8(8)}$ and
$SO(n+3,m+3)$, which are the quasiconformal groups of split
non-Euclidean Jordan algebras of degree three \footnote{ Actually, maximally split orthogonal groups correspond to the case $n=m$. For $n=1$ one gets the quasiconformal group of a Euclidean Jordan algebra.}. More specifically,
in section 2 we review the necessary background regarding Euclidean
and non-Euclidean Jordan algebras of degree three  and their rotation (
automorphism) and Lorentz (reduced structure) groups.  The U-duality
symmetries of maximal supergravity in five, four and three dimensions
are simply the Lorentz, conformal and quasiconformal groups of the
split exceptional Jordan algebra $J_3^{\mathbb{O}_S}$. The
corresponding symmetry groups of $N=4$ (sixteen supercharges)
Maxwell-Einstein-Supergravity theories are determined by the
non-simple Jordan algebras $\mathbb{R}\oplus \Gamma_{(5,n)}$. In
section 3 we review the conformal symmetry groups of the relevant
Jordan algebras.  In Section 4 we present the unified quasiconformal
realizations of $E_{6(6)}$, $E_{7(7)}$, $E_{8(8)}$ and $SO(n+3,m+3)$
twisted by a unitary character $\nu$ and their commutation relations as well as 
their quadratic Casimir operators. We determine the values of the
Casimir operators as a function of $\nu$ and the dimension $n_V$ of
the underlying Jordan algebra $J$. From this we determine the values
of $\nu$ for which the quasiconformal action induces unitary
representations on the space of square integrable functions in
$(2n_V+3)$ variables, that belong to the principle series. In section
5 we present the spherical vectors of these quasiconformal groups in a
unified manner and discuss each group separately. We also present the
compact spaces corresponding to the ``quasiconformal compactification''
of the $(2n_V+3)$ dimensional spaces on which the quasiconformal
groups $QConf(J)$ act for all Jordan algebras of degree three,
Euclidean as well as split. In section 6 we point out the connection
between real and split exceptional Jordan algebras of degree three and
the quaternionic Jordan algebras of degree four and discuss the
similarities and differences between the exceptional $N=2$
supergravity and maximal $N=8$ supergravity as they relate to
quaternionic Jordan algebras of rank four.

\renewcommand{\theequation}{\arabic{section}.\arabic{subsection}.\arabic{equation}}
\setcounter{equation}0
\section{ Euclidean (compact) and non-Euclidean (noncompact) Jordan algebras of Degree Three}
%\section{ Euclidean and Split Jordan algebras of Degree Three}
Referring to the monograph \cite{MR2014924} for details and
references on the subject we shall give a brief review of Jordan
algebras in this section, focussing mainly on Jordan algebras of
degree three.

A Jordan algebra over a field $\mathbb{F}$, which we take to generally
to be the real numbers $\mathbb{R}$, is an algebra, $J$ , with a
symmetric product $\circ$
\begin{equation}\label{commute} X\circ Y = Y
\circ X \in J, \quad \forall\,\, X,Y \in J \ ,
\end{equation}
such that the Jordan identity holds: 
\begin{equation}\label{Jidentity}
X\circ (Y \circ X^2)= (X\circ Y) \circ X^2 \ ,
\end{equation}
where $X^2\equiv (X\circ X)$.  Hence a Jordan algebra is commutative
and in general not associative algebra. They were introduced by
Pascual Jordan in his attempt to generalize the formalism of quantum
mechanics and finite dimensional simple Jordan algebras were
classified by him, von Neumann and Wigner \cite{Jordan:1933vh}.

A Jordan algebra $J$ is said to be Euclidean if for any two elements $X$ and $Y$ of $J$ the condition  
\begin{equation*}
   X\circ X + Y \circ Y =0
\end{equation*}
implies that both $X$ and $Y$ must vanish. Since the automorphism
groups of Euclidean Jordan algebras are compact they are also referred
to as compact. Otherwise the Jordan algebra is referred to as
noncompact or non-Euclidean. One can in general define a norm form,
$\mathbf{N}:J\rightarrow \mathbb{R}$ over $J$ that satisfies the
composition property \cite{MR0251099}
\begin{equation}
  \label{Norm}
  \mathbf{N}(\{X,Y,X \}) =\mathbf{N}^{2}(X) \mathbf{N}(Y)
\end{equation}
where $\{X,Y,Z\}$ is the Jordan triple product defined as
\begin{equation}
\{X,Y,Z\} = X\circ (Y\circ Z) + Z\circ (Y\circ X) - (X\circ Z) \circ Y 
\end{equation}
The degree, $p$, of the norm form as well as of $J$ is defined by the homogeneity condition 
$\mathbf{N}(\lambda X)=\lambda^p \mathbf{N}(X)$, where $\lambda\in \mathbb{R}$.  
%%%%%%%%%%%% %%%%%%%%%
\subsection{Euclidean Jordan Algebras of degree three and $5D$, $N=2$ MESGT's}
As was shown in \cite{Gunaydin:1983bi}, there exists a one-to-one
correspondence between Euclidean Jordan algebras of degree three and
the $5D$, $N=2$ MESGT's whose scalar manifolds are symmetric spaces
such that $G$ is a symmetry of their Lagrangian. In these theories the
symmetric C-tensor that describes the $F \wedge F \wedge A$ type
coupling \footnote{ We should note that a given $N=2$ MESGT in five
dimensions is uniquely determined by the C-tensor.}
\begin{equation*}
  C_{IJK} \epsilon_{\mu\nu\lambda\rho\sigma} F^{I\mu\nu} F^{J\lambda\rho} A^{J\sigma}
\end{equation*}
of all the vector fields including the graviphoton is identified with
the symmetric tensor that defines the cubic norm of the corresponding
Euclidean Jordan algebra $J$ of degree three.  Their scalar manifolds
are of the form
\begin{equation}
\mathcal{M}_5 (J) = \frac{Str_0(J)}{Aut(J)}
\end{equation}
where $Str_0(J)$ is the Lorentz (reduced structure) group of $J$ and
$Aut(J)$ is its rotation (automorphism) group. For Euclidean Jordan
algebras the rotation (automorphism) groups are compact.

There exists an infinite family of non-simple Jordan algebras of
degree three which are the direct sum of a one dimensional Jordan
algebra $\mathbb{R}$ and a Jordan algebra $\Gamma_{(1,n-1)}$
associated with a quadratic form of Lorentzian signature:
\begin{equation}
  J=\mathbb{R}\oplus \Gamma_{(1,n-1)}
\end{equation}
which is referred to as the generic Jordan family.  A simple
realization of $\Gamma_{(1,n-1)}$ is provided by $2^{[n/2]} \times
2^{[n/2]} $ Dirac gamma matrices $\gamma^i$
$(i,j,\ldots=1,\ldots,(n-1))$
%$(n-1)$ Dirac gamma matrices $\gamma^i$ $(i,j,\ldots=1,\ldots,(n-1))$
of an $(n-1)$ dimensional Euclidean space together with the identity
matrix $ \gamma^0 = \mathbf{1}$ and the Jordan product $\circ$ being
one half the anticommutator:

\begin{eqnarray}
\gamma^i \circ \gamma^j &=& \frac{1}{2}
\{\gamma^i,\gamma^j\}= \delta^{ij} \gamma^0
\nonumber \\
\gamma^0 \circ \gamma^0 &=& \frac{1}{2}
\{\gamma^0,\gamma^0\}= \gamma^0 \nonumber\\
\gamma^i \circ \gamma^0 &=& \frac{1}{2}
\{\gamma^i,\gamma^0\}= \gamma^i \ .
\end{eqnarray}
The quadratic norm of a general element $\mathbb{X} = X_0 \gamma^0 + X_i \gamma^i $ of
$\Gamma_{(1,n-1)}$
is defined as
\begin{equation*}
  \mathbf{Q}(\mathbb{X}) = \frac{1}{2^{[n/2]}} \mathop{Tr}
  \mathbb{X} \bar{\mathbb{X}} = X_0X_0 - X_iX_i \ ,
\end{equation*}
where
\begin{equation*}  \bar{\mathbb{X}} \equiv  X_0 \gamma^0 - X_i \gamma^i  \ .
\end{equation*}
The norm of a general element $\xi \oplus \mathbb{X} $ of the non-simple
Jordan algebra $J=\mathbb{R}\oplus \Gamma_{(1,n-1)}$ is
simply given by
\begin{equation}
  \mathbf{N}(\xi \oplus \mathbb{X}) =  \xi \mathbf{Q}(\mathbb{X})
\end{equation}
where $\xi \in \mathbb{R}$.

The scalar manifolds of corresponding 5D, $N=2$ MESGT's are
\begin{equation*}
 \mathcal{M} =\frac{SO(n -1,1)}{SO(n -1)}\times  SO(1,1)
\end{equation*}

In addition to the generic infinite family there exist four simple
Euclidean Jordan algebras of degree three.  They are generated by
Hermitian $(3\times 3)$-matrices over the four division algebras
$\mathbb{A} = \mathbb{R}, \mathbb{C}, \mathbb{H}, \mathbb{O}$
\begin{equation*}
 J =
\left(
  \begin{array}{ccc}
    \alpha & Z & \bar{Y} \\
    \bar{Z} & \beta & X \\
    Y & \bar{X} & \gamma \\
  \end{array}
\right)
\end{equation*}
where $\alpha , \beta, \gamma \in \mathbb{R}$ and $X,Y,Z \in
\mathbb{A}$ with the product being one half the anticommutator.  They
are denoted as $J_3^{\mathbb{R}}$, $J_3^{\mathbb{C}}$,
$J_3^{\mathbb{H}}$, $J_3^{\mathbb{O}}$, respectively, and the
corresponding $N=2$ MESGts are called ``magical supergravity theories''.  
They have the 5D scalar manifolds:
\begin{eqnarray}\label{magicals}
J_{3}^{\mathbb{R}}:\quad \mathcal{M}&=& SL(3,\mathbb{R})/
SO(3)\qquad
\nonumber\\
J_{3}^{\mathbb{C}}:\quad \mathcal{M}&=& SL(3,\mathbb{C})/
SU(3)\qquad
\nonumber\\
J_{3}^{\mathbb{H}}:\quad \mathcal{M}&=& SU^{*}(6)/
USp(6)\qquad
 \nonumber \\
J_{3}^{\mathbb{O}}:\quad \mathcal{M}&=& E_{6(-26)}/
F_{4}\qquad \qquad
\end{eqnarray}

The cubic norm form, $\mathbf{N}$, of the simple Jordan algebras of
degree three is given by the ``determinant'' of the corresponding
Hermitian $(3\times 3)$-matrices (modulo an overall scaling factor).
\begin{equation}
\mathbf{N}(J)= \alpha \beta \gamma - \alpha X \bar{X} - \beta Y \bar{Y} - \gamma Z \bar{Z} + 2 Re (X Y Z )
\end{equation}
where $Re(XYZ)$ denotes the real part of $XYZ$ and bar denotes
conjugation in the underlying division algebra.

For a real quaternion $X \in \mathbb{H}$ we have
\begin{eqnarray}
X&=&X_0 + X_1 j_1 + X_2 j_2 + X_3 j_3 \nonumber \\
\bar{X}& = &X_0 - X_1 j_1 - X_2 j_2 - X_3 j_3 \\
X \bar{X}& =& X_0^2 + X_1^2 + X_2^2 +X_3^2 \nonumber
\end{eqnarray}
where the imaginary units $j_i$ satisfy
\begin{equation}
j_i j_j = - \delta_{ij} + \epsilon_{ijk} j_k
\end{equation}
For a real  octonion $X\in \mathbb{O}$ we have
\begin{eqnarray}
X &=& X_0 + X_1 j_1 + X_2 j_2 + X_3 j_3  + X_4 j_4 + X_5 j_5  +X_6 j_6 + X_7 j_7 \nonumber \\
\bar{X} &=& X_0 - X_1 j_1 - X_2 j_2 - X_3 j_3  - X_4 j_4 - X_5 j_5  - X_6 j_6 - X_7 j_7 \\
X \bar{X} &= &X_0^2 + \sum_{A=1}^7 (X_A)^2 \nonumber
\end{eqnarray}
Seven imaginary units of real  octonions satisfy
\begin{equation}
j_A j_B = - \delta_{AB} + \eta_{ABC} j_C
\end{equation}
where $\eta_{ABC}$ is completely antisymmetric and in the conventions of \cite{Gunaydin:1973rs} take on the values
\begin{equation}
\eta_{ABC} = 1  \Leftrightarrow  (ABC)= (123), (471), (572), (673), (624), (435), (516)
\end{equation}
The automorphism group of the division algebra of octonions is the compact group $G_2$. \\

\FIGURE[r]{
  \includegraphics{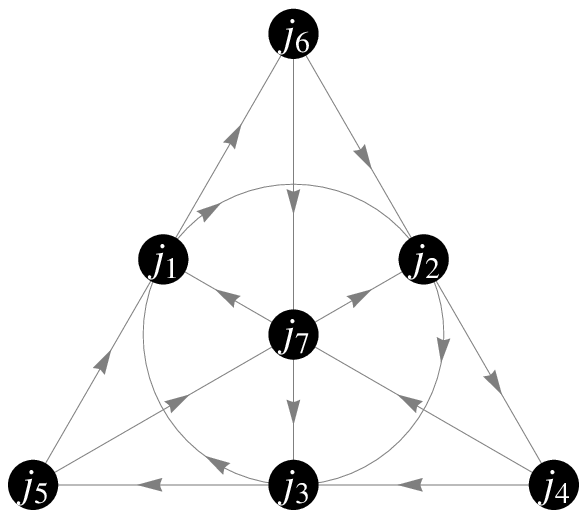}
  \caption{Multiplication table of imaginary units of real octonions $\mathbb{O}$. Three imaginary units on each side, height and circle correspond to the imaginary units of a quaternion subalgebra. The arrows represent the positive directions for multiplication, e.g. $j_1 j_2 =-j_2 j_1=j_3 $ and $j_6j_2=-j_2j_6=j_4$, etc..}
}

We should note that the simple Jordan algebras $J_3^{\mathbb{A}}$ of
degree 3 have nonsimple subalgebras generated by elements of the form
\begin{equation*}
 J =
\left(
  \begin{array}{ccc}
    \alpha & 0 & 0 \\
    0 & \beta & X \\
    0 & \bar{X} & \gamma \\
  \end{array}
\right)
\end{equation*}
that are isomorphic to generic Jordan algebras $(\mathbb{R}\oplus
\Gamma_{(1,2)})$, $(\mathbb{R}\oplus \Gamma_{(1,3)})$, 
$(\mathbb{R}\oplus \Gamma_{(1,5)})$ and $(\mathbb{R}\oplus \Gamma_{(1,9)})$ 
for $\mathbb{A}= \mathbb{R}, \mathbb{C}, \mathbb{H}$ and $\mathbb{O}$, respectively.

\subsection{ Non-Euclidean Jordan algebras of degree three and $5D$, $N \geqslant 4 $ Supergravity Theories}

In the generic infinite family of non-simple Jordan algebras of degree
three, $\mathbb{R} \oplus \Gamma$, one can take the quadratic form
defining the Jordan algebra $\Gamma$ of degree two to be of arbitrary
signature different from Minkowskian, which result in non-compact or
non-Euclidean Jordan algebras. If the quadratic norm form has
signature $(n,m)$ we shall denote the Jordan algebra as
$\Gamma_{(n,m)}$.
$\Gamma_{(n,m)}$ is realized by $2^{[(n+m)/2]} \times 2^{[(n+m)/2]} $
Dirac gamma matrices $\gamma^i$ $(i,j,\ldots=1,\ldots,(n+m-1))$
together with the identity matrix $\gamma^0 = \mathbf{1}$ and the
Jordan product $\circ$ being one half the anticommutator:

\begin{eqnarray}
\gamma^i \circ \gamma^j &=& \frac{1}{2}
\{\gamma^i,\gamma^j\}= \eta^{ij} \gamma^0
\nonumber \\
\gamma^0 \circ \gamma^0 &=& \frac{1}{2}
\{\gamma^0,\gamma^0\}= \gamma^0 \nonumber\\
\gamma^i \circ \gamma^0 &=& \frac{1}{2}
\{\gamma^i,\gamma^0\}= \gamma^i \ .
\end{eqnarray}
where
\begin{eqnarray}
\eta^{ij} &= &\delta^{ij} \,\,\, , \,\, for \,\,\, i,j,..=1,2,..,(n-1) \nonumber \\
\eta^{ij} &=& - \delta^{ij} \,\,\, , \,\, for \,\,\,  i,j,..=n,n+1,..,(n+m-1) 
\end{eqnarray}
The quadratic norm of a general element $\mathbb{X} = X_0 \gamma^0 +
X_i \gamma^i $ of $\Gamma_{(n,m)}$ is given by
\begin{equation*}
  \mathbf{Q}(\mathbb{X}) = \frac{1}{2^{[n/2]}} \mathop{Tr}
  \mathbb{X} \bar{\mathbb{X}} = X_0X_0 + \eta^{ij} X_iX_j \ ,
\end{equation*}
where
\begin{equation*}  
  \bar{\mathbb{X}} \equiv  X_0 \gamma^0 - X_i \gamma^i  \ .
\end{equation*}

The norm of a general element $\xi \oplus \mathbb{X} $ of 
non-simple Jordan algebra $J=\mathbb{R}\oplus \Gamma_{(n,m)}$ is
simply given by
\begin{equation}
  \mathbf{N}(\xi \oplus \mathbb{X}) =  \xi \mathbf{Q}(\mathbb{X})
\end{equation}
where $\xi \in \mathbb{R}$. The invariance group of the cubic norm is
\begin{equation}
Str_0(\mathbb{R}\oplus \Gamma_{(n,m)}) = SO(1,1)\times SO(n,m)
\end{equation}

If one replaces the underlying division algebra $\mathbb{A}$ of three
of the four simple Euclidean Jordan algebras $J_3^{\mathbb{A}}$ by the
corresponding split composition algebras $\mathbb{A}_S$ one obtains
non-Euclidean simple Jordan algebras $J_3^{\mathbb{A}_S}$ for
$\mathbb{A}_S = \mathbb{C}_S, \mathbb{H}_S $ and $ \mathbb{O}_S$.

For split octonions $\mathbb{O}_S$ , four of the seven ``imaginary
units'' square to $+1$, while the other three square to -1. If we
denote the split imaginary units as $j_{\mu}^s$ ($\mu =4,5,6,7$) and
the imaginary units of the real quaternion subalgebra as $j_i$, 
$(i=1,2,3)$ we have:
\begin{eqnarray}
j_{\mu}^s j_{\nu}^s& =& \delta_{\mu \nu} - \eta_{\mu \nu i} j_i \nonumber \\
j_i j_j &=& -\delta_{ij} + \epsilon_{ijk} j_k \\
j_i j^s_{\mu} &=& \eta_{i\mu\nu} j^s_\nu \nonumber
\end{eqnarray}
where $\eta_{ABC}$ $(A,B,C=1,2..7)$ are the structure constant of the
real octonion algebra $\mathbb{O}$ defined above.  For a split
octonion
\begin{equation*}
  O_s = o_0 + o_1 j_1 + o_2 j_2 + o_3 j_3  + o_4 j^s_4 + o_5 j^s_5  +o_6 j^s_6 + o_7 j_7^s
\end{equation*}
the norm is
\begin{equation*} 
  O_s \bar{O}_s = o_0^2 +o_1^2 + o_2^2 +o_3^2 - o_4^2 -o_5^2 -o_6^2 -o_7^2 
\end{equation*}
where $\bar{O}_s = o_0 - o_1 j_1 - o_2 j_2 - o_3 j_3 - o_4 j^s_4 - o_5
j^s_5 - o_6 j^s_6 - o_7 j_7^s $. The norm has the invariance group
$SO(4,4)$. The automorphism group of split octonions is the
exceptional group $G_{2(2)}$ with the maximal compact subgroup
$SU(2)\times SU(2)$.

The automorphism group of the split exceptional Jordan algebra defined
by $ 3\times 3 $ split octonionic Hermitian matrices of the form
\begin{equation}
J^s =\left(
       \begin{array}{ccc}
         \alpha & Z^s& \bar{Y}^s\\
         \bar{Z}^s & \beta & X^s \\
         Y^s & \bar{X}^s & \gamma \\
       \end{array}
     \right)
\end{equation}
is the noncompact group $F_{4(4)}$ with the maximal compact subgroup
$USp(6)\times SU(2)$ and its reduced structure group is $E_{6(6)}$
with the maximal compact subgroup $USp(8)$.

The split exceptional Jordan algebra has a subalgebra generated by
elements of the form
\begin{equation}
J^s =\left(
       \begin{array}{ccc}
         \alpha & 0& 0\\
         0 & \beta & X^s \\
         0 & \bar{X}^s & \gamma \\
       \end{array}
     \right)
\end{equation}
which is isomorphic to $(\mathbb{R}+ \Gamma_{(5,5)})$, whose reduced
structure and automorphism groups are $SO(5,5)\times SO(1,1)$ and
$SO(4,5)$, respectively.

The split quaternion algebra $\mathbb{H}^s$ has two ``imaginary
units'' $j^s_m$ ($m=2,3$) that square to +1:
\begin{eqnarray}
j^s_m j^s_n = \delta_{mn} - \epsilon_{mnk} j_k \\
(j_1)^2 =-1 \nonumber \\
j_1 j^s_m = \epsilon_{1mn} j_n^s \nonumber
\end{eqnarray}
For a split quaternion
\begin{equation*}
  Q_s = q_0 + q_1 j_1 + q_2 j_2^s + q_3 j^s_3
\end{equation*}
the norm is
\begin{equation*}
     Q_s \bar{Q}_s = q_0^2 +q_1^2 -q_2^2 -q_3^2
\end{equation*}
where $\bar{Q}_s = q_0 - q_1 j_1 -q_2 j_2^s -q_3 j^s_3 $ and it is
invariant under $SO(2,2)$.  The automorphism group of the split Jordan
algebra $J_3^{\mathbb{H}_S}$ is $Sp(6,\mathbb{R})$ with the maximal
compact subgroup $SU(3)\times U(1)$. Its reduced structure group is
$SL(6,\mathbb{R})$.

The split complex numbers have an ``imaginary unit'' that squares to +1
and the norm has $SO(1,1)$ invariance. The automorphism group of split
complex Jordan algebra $J_3^{\mathbb{C}_s}$ is $SL(3,\mathbb{R})$ and
its reduced structure group is
\begin{equation*}
  SL(3,\mathbb{R})\times SL(3,\mathbb{R})
\end{equation*}
%\appendix

The invariant tensor $C_{IJK}$ defining the cubic norm of the Jordan algebra
$(\mathbb{R}\oplus \Gamma_{(5,n)})$ can be identified with the
C-tensor in the  $F \wedge
F \wedge A $ coupling 
\begin{equation}
C_{IJK} F^I \wedge F^J \wedge A^K = C_{IJK} \epsilon_{\mu\nu\lambda\sigma\rho} F^{I\mu\nu} F^{J\lambda\sigma} A^{K\rho} 
\end{equation}
of $(n+5)$ vector fields of $N=4$ MESGT's that describe the coupling of
$n$ vector multiplets to $N=4$ supergravity in five
dimensions. Scalar manifolds of these theories are symmetric spaces
\begin{equation}
\mathcal{M}_5= \frac{ SO(5,n)\times SO(1,1) }{SO(5) \times SO(n)}
\end{equation}
where the isometry group $SO(5,n)\times SO(1,1)$ is simply the
reduced structure group of the Jordan algebra
$(\mathbb{R}\oplus\Gamma_{(5,n)})$.

The bosonic field content of $N=6$ simple supergravity is the same as
that of $N=2$ MESGT defined by the Euclidean Jordan algebra
$J_3^{\mathbb{H}}$ \cite{Gunaydin:1983rk}. Hence the scalar manifold
of $N=6$ supergravity is
\begin{equation} 
   \mathcal{M}_5(N=6) = \frac{Str_0(J_3^{\mathbb{H}})}{Aut(J_3^{\mathbb{H}})}=\frac{SU^*(6)}{USp(6)} 
\end{equation}
Therefore its invariant C-tensor is simply the one given by the cubic
norm of $J_3^{\mathbb{H}}$, which is a Euclidean Jordan algebra.

As for N=8 supergravity in five dimensions its C-tensor is simply the
one given by the cubic norm of the split exceptional Jordan algebra
$J_3^{\mathbb{O}_S}$ defined over split octonions $\mathbb{O}_S$.
$E_{6(6)}$ is the invariance group of the C-tensor as well as of the
full maximal supergravity in five dimensions whose scalar manifold is
\begin{equation} 
  \mathcal{M}_5 (N=8) =\frac{E_{6(6)}}{USp(8)} 
\end{equation}

The five dimensional $N=8$ supergravity can be truncated to $N=4$ MESGT describing the coupling of  5 vector multiplets to $N=4$ supergravity with the scalar manifold
\begin{equation}
\frac{SO(1,1)\times SO(5,5)}{SO(5)\times SO(5)}
\end{equation}
The C-tensor of this truncated theory coincides with the symmetric tensor that
defines the cubic norm of the Jordan subalgebra $(\mathbb{R}\oplus
\Gamma_{(5,5)})$ of $J_3^{\mathbb{O}_S}$ given above.

\section{Conformal Groups of Jordan Algebras and U-duality groups in $d=4$ }
\renewcommand{\theequation}{\arabic{section}.\arabic{equation}}
\setcounter{equation}{0} The proposal \cite{Gunaydin:1975mp} to
define generalized spacetimes coordinatized by elements of a Jordan
algebra $J$ identifies the automorphism $Aut(J)$ and reduced structure
groups $Str_0(J)$ of the Jordan algebra $J$ with the rotation and
Lorentz groups of the corresponding spacetime respectively. The
Lorentz group and dilatations generate the structure group $Str(J)$ of
$J$, which extends to the generalized conformal group $Conf(J)$ of
the Jordan algebra $J$. $Conf(J)$  can be identified with the invariance
group of the light cone defined by the norm form of the Jordan algebra $J$
\cite{Gunaydin:1975mp,Gunaydin:1989dq,Gunaydin:1992zh,Gunaydin:2000xr}. For
Euclidean Jordan algebras of dimensions $n $ defined by a quadratic
norm form with a Lorentzian signature the rotation, Lorentz and
conformal groups are simply $SO(n-1)$, $SO(n-1,1)$ and $SO(n,2)$,
respectively.
 
Lie algebra $\mathfrak{conf}(J)$ of the generalized conformal group
$Conf(J)$ has a natural 3-grading with respect to the dilatation
generator $\mathcal{R}$. Choosing a basis $e_I$ for the Jordan
algebra and labelling the translations and special conformal
generators as $T_I$ and $K^I$, respectively, we have
\begin{equation}
 \mathfrak{conf}(J) = T_I \oplus R_I^J \oplus K^I
\end{equation}
where $I,J,..=1,2,...,dim(J)=n_V$.  Traceless components  $L_I^J$ of $R_I^J$
are the Lorentz group generators and the trace part is proportional to
the dilatation generator $\mathcal{R}$ :
\begin{equation*}
 \mathcal{R}= \frac{1}{n_V} R_K^K
\end{equation*}
\begin{equation}
 R_I^J = L_I^J + \delta_I^J \mathcal{R}
\end{equation}

In the chosen basis $e_I$ for the Jordan algebra an element
$\mathbf{x} \in J$ can be written as $\mathbf{x} = e_I q^I=\tilde{e}^I
q_I$. Then the action of the generators of $\mathfrak{conf}(J)$ on $J$ can
be written as differential operators acting on the ``coordinates''
$q^I$.\footnote{ Note that there are , in general, two inequivalent
actions of the reduced structure group: one on Jordan algebra and
another one on its conjugate (or dual). The tilde refers to the
conjugate basis such that $q_I p^I$ is invariant under the action of
reduced structure group $Str_0(J)$. For details on this issue see
\cite{Gunaydin:1992zh}.} These generators can be twisted by a unitary
character $\lambda$ and take the simple form
\begin{eqnarray}
  T_I & = &\frac{\partial}{\partial q^I}  \nonumber \\
  R^I_J& = &- \Lambda^{IK}_{JL} q^L \frac{\partial}{\partial q^K}  - \lambda  \delta^I_J \\
  K^I & =& \frac{1}{2} \Lambda^{IK}_{JL}  q^J q^L  \frac{\partial}{\partial q^K} + \lambda q^I \nonumber \\
\end{eqnarray}

They satisfy the commutation relations
\begin{eqnarray}
&[ T_I, K^J ] =& - R^J_I \\
&[ R^J_I , T_K ] = &\Lambda_{IK}^{JL} T_L \\
&[R^J_I , K^K ] = &- \Lambda_{IK}^{JL} K^L
\end{eqnarray}
where $\Lambda_{KL}^{IJ} $ are the structure constants of the Jordan triple product
\begin{equation}
\{e_I, \tilde{e}^K , e_J \} =\Lambda^{KL}_{IJ} e_L
\end{equation}
\begin{equation}
\{ \tilde{e}^K,e_I,\tilde{e}^L\} = \Lambda^{KL}_{IJ} \tilde{e}^J
\end{equation}
The generators  of rotation (automorphism) group $Aut(J)$  are
\begin{equation}
A_{IJ} = R_I^J -R_J^I
\end{equation}
For Jordan algebras of degree three the structure constants can be written as  
\begin{equation}
   \Lambda_{KL}^{IJ} := \delta_K^I \delta_L^J + \delta_L^I \delta^J_K - \frac{4}{3} C^{IJM} C_{KLM}
\end{equation}
where $C_{IJK}$ is the symmetric tensor that defines the cubic norm of
$J$ and satisfies the ``adjoint identity''~\cite{Gunaydin:1983bi}:
\begin{equation}
   C^{IJK} C_{J(MN} C_{PQ)K} = \delta^{I}_{(M} C_{NPQ)}
\end{equation}
Since the C-tensor is an invariant of the Lorentz (reduced structure)
group we have
\begin{equation}
C_{IJK}=C^{IJK} 
\end{equation}

The conformal groups of  non-Euclidean Jordan algebras
$J_3^{\mathbb{C}_s},J_3^{\mathbb{H}_s}, J_3^{\mathbb{O}_s}$ and
$(\mathbb{R}\oplus \Gamma_{(m,n)}) $ are listed in
Table\ref{noneuclideantheories}. The conformal group of $
J_3^{\mathbb{O}_s}$ is $E_{7(7)}$ which is the U-duality group of
maximal supergravity in $d=4$. Similarly, the conformal groups
$SO(6,n+1) \times SU(1,1)$ of the Jordan algebras $(\mathbb{R}\oplus
\Gamma_{(5,n)}) $ are the U-duality groups of $N=4$ MESGTs in $d=4$.

\begin{small}
\begin{table}
\begin{equation} \nonumber
  \begin{array}{|c|c|c|c|c|} \hline
  J & Aut(J) & Str_0(J) & Conf(J) & QConf(J) \\ \hline
  J_3^{\mathbb{C}_s} & SL(3,\mathbb{R}) & SL(3,\mathbb{R}) \times SL(3,\mathbb{R}) & SL(6,\mathbb{R}) & E_{6(6)} \\ \hline
  J_3^{\mathbb{H}_s} & Sp(6,\mathbb{R}) & SL(6,\mathbb{R}) & SO(6,6) &E_{7(7)} \\ \hline
  J_3^{\mathbb{O}_s} & F_{4(4)} & E_{6(6)} & E_{7(7)} & E_{8(8)} \\ \hline
\mathbb{R} \oplus \Gamma_{(n,m)} & SO(n-1,m) & SO(1,1)\times SO(n,m) & SO(n+1,m+1) \times SU(1,1) & SO(n+3,3+m) \\ \hline
\end{array}
\end{equation}
\caption{\label{noneuclideantheories} 
Above we give the automorphism
($Aut(J)$), reduced structure $Str_0(J)$, conformal ($Conf(J)$) and
quasiconformal groups $(QConf(J))$ associated with non-Euclidean
Jordan algebras of degree three.  }
\end{table}
\end{small}

\section{Quasiconformal Groups associated with  Non-Euclidean Jordan Algebras of Degree Three}
%
%\renewcommand{\theequation}{\arabic{section}.\arabic{equation}}
%
%\subsection{Quasiconformal Groups as Invariance Groups of  Quartic Light-cones}
\renewcommand{\theequation}{\arabic{section}.{\arabic{subsection}}.\arabic{equation}}
\setcounter{equation}{0}
Quasiconformal realizations of Lie groups was first formulated over
Freudenthal triple systems associated with Lie groups  extended by an extra singlet coordinate 
 \cite{Gunaydin:2000xr}.  Given a simple Lie  algebra $ \mathfrak{g}$ one can associate a  Freudenthal triple system $\mathcal{F}$  with it  via the Freudenthal-Kantor construction \cite{kansko} using the 5-grading of  $\mathfrak{g}$
\begin{equation}
  \mathfrak{g}= g^{-2}\oplus g^{-1} \oplus g^0 \oplus g^{+1} \oplus g^{+2} 
\end{equation}
such that grade $\pm 2$ subspaces are one dimensional. The generators
of grade $+1$ and by conjugation also of $-1$ subspaces are labelled
by the elements of the underlying FTS:
\begin{equation}
\mathfrak{g}= \Tilde{K} \oplus \Tilde{U}_A \oplus  S_{AB} \oplus U_B \oplus K
\end{equation}
where $A,B \in \mathcal{F}$.  Here we shall focus on quasiconformal
realizations of groups associated with FTS's defined by non-Euclidean
Jordan algebras $J$ of degree three in a basis covariant with respect
to their Lorentz (reduced structure) groups.  The elements of a FTS
$\mathcal{F}(J)$ defined over $J$ can be represented as formal 
$2 \times 2$ ``matrices''
\begin{equation} X= 
    \begin{pmatrix}
        \alpha & \mathbf{x} \cr
    \mathbf{y} & \beta
    \end{pmatrix} 
\end{equation}
where $\alpha, \beta \in \mathbb{R}$ and $\mathbf{x}, \mathbf{y} \in
J$. We shall write $X$ simply as $X \equiv \left(\alpha, \beta,
\mathbf{x}, \mathbf{y}\right)$ for convenience. Every FTS admits a
skew-symmetric bilinear form. Given two elements
$X=\left(\alpha,\beta,\mathbf{x},\mathbf{y}\right)$ and
$Y=\left(\gamma,\delta,\mathbf{w},\mathbf{z}\right)$ of
$\mathcal{F}(J)$ their skew symmetric bilinear form is
\begin{equation}
  \left< X, Y\right> \equiv
         \alpha \delta - \beta\gamma +
         \left(\mathbf{x}, \mathbf{z}\right) -
     \left(\mathbf{y}, \mathbf{w}\right)
\end{equation}
where $\left(\mathbf{x}, \mathbf{z}\right)$ is the symmetric bilinear
form over $J$ given by the trace $\mathrm{Tr}$ :
\begin{equation}
    \left( \mathbf{x}, \mathbf{z}\right) \equiv \mathop\mathrm{Tr}\left( \mathbf{x} \circ \mathbf{z}\right)
\end{equation}

In the normalization and conventions of \cite{Gunaydin:2000xr} the
quartic norm $\mathcal{Q}_4 \left(X\right)$ of
$X \in \mathcal{F}(J)$ is given by
\begin{equation}
\mathcal{Q}_4 \left(X\right) \equiv \frac{1}{48} \left< \left(X, X, X\right), X \right>
\end{equation}
where $\left(X, Y, Z\right)$ denotes the Freudenthal triple product.
The automorphism group of the FTS $\mathcal{F}(J)$ is isomorphic to the conformal group of the Jordan algebra $J$:
\begin{equation}
Aut(\mathcal{F}(J)) \cong Conf(J)
\end{equation}
We note that under the action of $Aut(\mathcal{F}(J))$ the elements
$X=\left(\alpha,\beta,\mathbf{x},\mathbf{y}\right)\in\mathcal{F}(J)$
transform linearly, which is not to be confused with the nonlinear
action of the conformal group $Conf(J)$ on $J$.  Under the Lorentz
subgroup $Str_0(J)$ of $Aut(\mathcal{F}(J))$ the Jordan components $
\mathbf{x}$ and $\mathbf{y}$ of $X$ transform in conjugate (dual)
representations.

For $5d $ supergravity theories whose C-tensors are
given by the norm forms of  Jordan algebras $J$ of degree three
one-to-one correspondence between the vector fields and the elements
of $J$ gets extended to a one-to-one correspondence between the vector
field strengths and their magnetic duals and elements of the
Freudenthal triple system $\mathcal{F}(J)$:
\begin{equation*}
(A^I_\mu \leftrightarrow J ) \,\,\, \Longrightarrow \Big( \begin{pmatrix}
        F^0 & F^I \cr
    \Tilde{F}^I  & \Tilde{F}^0 
    \end{pmatrix}  \,\,\, \rightarrow \mathcal{F}(J)  \Big)
\end{equation*}
where $F^0$ denotes the field strength of the vector field that comes
from the 5d graviton.  Field strengths $F^I$ and their magnetic duals
$\Tilde{F}^I$ transform in conjugate representations under the Lorentz
group $Str_0(J)$. Since the automorphism group of a Freudenthal triple
system $\mathcal{F}(J)$ defined over a Jordan algebra $J$ of degree
three is isomorphic to the four dimensional U-duality group
$Aut(\mathcal{F}(J))=Conf(J) $ of corresponding supergravity theories
the original formulation of \cite{Gunaydin:2000xr} is covariant with
respect to $Conf(J)$.
 
Consider now the vector space $\mathcal{T}$ of a FTS extended by an extra
coordinate. We shall denote vectors in this space as $\cX=(X,x) \in
\mathcal{T}$ where $X$ belongs to the FTS and $x$ is the extra
coordinate.  The action of Lie algebra of the quasiconformal group,
associated with a FTS $\mathcal{F}$, on the vector space $\mathcal{T}$
 is given by
\cite{Gunaydin:2000xr,Gunaydin:2005zz}:
\begin{equation}
\begin{split}
  \begin{aligned}
      K\left(X\right) &= 0 \\
      K\left(x\right) &= 2\,
  \end{aligned}
  & \quad
  \begin{aligned}
     U_A \left(X\right) &= A \\
     U_A\left(x\right) &= \left< A, X\right>
  \end{aligned}
   \quad
   \begin{aligned}
      S_{AB}\left(X\right) &= \left( A, B, X\right) \\
      S_{AB}\left(x\right) &= 2 \left< A, B\right> x
   \end{aligned}
 \\ \label{qcg}
 &\begin{aligned}
    \Tilde{U}_A\left(X\right) &= \frac{1}{2} \left(X, A, X\right) - A x \\
    \Tilde{U}_A\left(x\right) &= -\frac{1}{6} \left< \left(X, X, X\right), A \right> + \left< X, A\right> x
 \end{aligned}
 \\
 &\begin{aligned}
    \Tilde{K}\left(X\right) &= -\frac{1}{6} \,  \left(X,X,X\right) +  X x \\
    \Tilde{K}\left(x\right) &= \frac{1}{6} \,  \left< \left(X, X, X\right), X \right> + 2\,  \, x^2
 \end{aligned}
\end{split}
\end{equation}
where $A,B \in \mathcal{F}$.

 The  quartic norm over the space
$\mathcal{T}$ is defined as
\begin{equation}
\cN_4(\cX) := \mathcal{Q}_4(X) - x^2
\end{equation}
where $\mathcal{Q}_4(X)$ is the quartic invariant of $X \in
\mathcal{F}$.  Quartic ``symplectic distance'' $d(\cX,\cY)$ between
any two points $\cX=(X,x)$ and $\cY=(Y,y) $ in $\mathcal{T}$ is
defined as the quartic norm of ``symplectic difference''
\begin{equation}
 \gd(\cX,\cY):=
(X-Y,x-y+\langle X, Y \rangle ) 
\end{equation}
of two vectors  in $\mathcal{T}$ 
\begin{equation}
  d(\cX,\cY):= \cN_4(\gd(\cX,\cY) = \mathcal{Q}_4 (X-Y) - \left( x-y+\langle X,Y \rangle \right)^2
  \label{quarticlightcone}
\end{equation}

 The quasiconformal group action defined above leaves invariant
light-like separations \cite{Gunaydin:2000xr}
\begin{equation}
  d(\cX,\cY)=0 \label{lightlike}
\end{equation}
In other words the quasiconformal group is the invariance group of
the light-cone with respect to the quartic distance function
\ref{quarticlightcone}.  We shall refer to the submanifold with base
point $\cX$ in the space $\mathcal{T}$ defined by the condition
\ref{lightlike} as the ``quartic light-cone''\footnote{By an abuse of
terminology we shall sometimes refer to the distance function
\ref{quarticlightcone} also as the quartic
light-cone.}. Quasiconformal realization of a simple Lie algebra
$\mathfrak{g}$ over a FTS $\mathcal{F}$ extended by an extra singlet
coordinate carries over to the complexification of
$\mathfrak{g}$. Therefore by taking different real sections one can
obtain quasiconformal realizations of different real forms of a the
corresponding group $G$.

The quartic light-cone \ref{quarticlightcone} is manifestly invariant
under the Heisenberg symmetry group corresponding to ``symplectic
translations'' generated by $U_A$ and $K$ in \ref{qcg}.  ``Symplectic
special conformal generators'' $\Tilde{U}_A$ and $\Tilde{K}$ also form
an Heisenberg subalgebra and their action on the quartic light-cone $
d\left(\cX, \cY\right)$
results in overall multiplicative factors
\cite{Gunaydin:2000xr,Gunaydin:2007qq}
\begin{equation}
 d\left(\cX,\cY \right)
  \Longrightarrow f(\cX,\cY ) \,\,  d\left(\cX,\cY \right)
\end{equation}
which proves that  light-like separations are left invariant under
the full quasiconformal group action.

For supergravity theories whose 5 and 4 dimensional U-duality symmetry
groups are the Lorentz and Conformal groups of a Jordan algebra $J$ of
degree three the  U-duality groups of corresponding 3d supergravity theories are isomorphic to
the quasiconformal groups $QConf(J)$ of the Jordan algebras $J$.

\subsection{Quasiconformal Lie Algebras of Non-Euclidean Jordan algebras of Degree Three  Twisted by a Unitary Character}
\renewcommand{\theequation}{\arabic{section}.{\arabic{subsection}}.\arabic{equation}}
\setcounter{equation}{0}
 
We shall denote the  basis vectors of $\mathcal{F}(J)$ as follows
\begin{equation}
 \begin{pmatrix} \alpha & \mathbf{x} \cr \mathbf{y} & \beta \end{pmatrix} = \alpha \mathbf{e}_0 + \beta \tilde{\mathbf{e}}^0
      + x^I \mathbf{e}_I + y_I \tilde{\mathbf{e}}^I
\end{equation}
where $I=1, \dots, n_V= dim(J)$ and $\mathbf{x}$ and $\mathbf{y}$
transform in conjugate representations of the Lorentz group
$Str_0(J)$.

The quasiconformal Lie algebra associated with a Jordan algebra $J$ of
degree three which we denote interchangeably as
$QConf(\mathcal{F}(J))$ or as $QConf(J)$ can be given a 7 by 5 graded
decomposition that is covariant with respect to the reduced structure
group $\mathop\mathrm{Str}_0(J) $ as shown in Table \ref{7by5grading}.
\begin{table} 
%\begin{equation}
\[
\begin{array}{ccccccc}
 \phantom{U}~~~ & \phantom{R}~~~ & \phantom{U}~~~ &  K ~~~ & \phantom{V}~~~ & \phantom{R}~~~ & \phantom{V}~~~ \\[4pt]
  U_0 & \phantom{R} & U_I & \vline \phantom{K} & V^I & \phantom{R} & V^0 \\[4pt]
 --- &  \Tilde{R}^I  &  --- & \left(  \mathcal{D} \oplus L_I^J \oplus \mathcal{R}  \right) &
             --- & R_J &  --- \\[4pt]
  \Tilde{U}_0 & \phantom{R} & \Tilde{U}_I & \vline \phantom{K} & \Tilde{V}^I & \phantom{R} & \Tilde{V}^0 \\[4pt]
  \phantom{U} & \phantom{R} & \phantom{U} &  \Tilde{K} & \phantom{V} & \phantom{R} & \phantom{V}
\end{array}
%\end{equation}
\]
\caption{\label{7by5grading} Above we give the $7 \times 5$ grading of
the quasiconformal Lie algebra $QConf(J)$ associated with the
Freudenthal triple system $\mathcal{F}(J)$ defined over a Jordan
algebra $J$ of degree 3. The vertical 5-grading is determined by
$\mathcal{D}=-\Delta$ that commutes with the Lorentz group generators
$L_I^J$ and with $\mathcal{R}$.  Horizontal 7-grading is determined by
$\mathcal{R}$.  The generators $\Tilde{R}^I, R_I^J $ and $R_I$
generate the automorphism group $Aut(\mathcal{F}(J))$ under which the
generators ($U_0,U_I, V^I, V^0 $) as well as ($ \Tilde{U}_0,
\Tilde{U}_I , \Tilde{V}^I , \Tilde{V}^0 $) transform linearly in a
symplectic representation.}
\end{table}
With applications to supergravity theories in mind we shall label the
elements $X,Y,..$ of FTS $\mathcal{F}(J)$ in terms of coordinates
($q_0$, $q_I$) and momenta ($p^0$, $p^I$) as follows\footnote{$p^I
q_I = \left(\mathbf{e}_I p^I, q_I \Tilde{\mathbf{e}}^I \right)$ and 
$\left(p^\sharp\right)_I \left(q^\sharp\right)^I = \left( {p^\sharp}_I
\Tilde{\mathbf{e}}^I, {q^\sharp}^I \mathbf{e}_I \right)$. }
\begin{equation}
    X = q_0 \Tilde{\mathbf{e}}^0 + q_I \Tilde{\mathbf{e}}^I + p^I \mathbf{e}_I + p^0 \mathbf{e}_0
\end{equation}
We shall normalize the basis elements and cubic norm (C-tensor) such
that the quartic invariant is given by
\begin{eqnarray}
I_4(X)& = &
    \left(p^0 q_0 -  p^I q_I \right)^2 - \frac{4}{3}  C_{IJK} p^J p^K C^{ILM} q_L q_M \\  \nonumber && +
             \frac{4}{3\sqrt{3}} p^0 C^{IJK} q_I q_J q_K + \frac{4}{3\sqrt{3}} q_0 C_{IJK} p^I p^J p^K \\ \nonumber
             &=& \left(p^0 q_0 -  p^I q_I \right)^2 - \frac{4}{3} (p^\sharp)_I  (q^\sharp)^I \\ && +
             \frac{4}{3\sqrt{3}} p^0 \mathcal{N}(q) + \frac{4}{3\sqrt{3}} q_0 \mathcal{N}(p)
         \nonumber
\end{eqnarray}
where
\begin{equation*}
 \begin{array}{ccc}
   \mathcal{N}(q) \equiv C^{IJK} q_I q_J q_K  & \phantom{seven letters}
   &(q^\sharp)^I   \equiv C^{IJK} q_J q_K \\
   \mathcal{N}(p) \equiv C_{IJK} p^I p^J p^K  &  &(p^\sharp)_I \equiv C_{IJK} p^J p^K
 \end{array}
\end{equation*}
The basis vectors $\mathbf{e}_I$ 
($\Tilde{\mathbf{e}}^I$) of the Jordan algebra $J$ (and its conjugate
$\Tilde{J}$) are normalized such that 
\begin{eqnarray}
    \left(\Tilde{e}^J, \Tilde{e}^I  \right)& =&  \mathop\mathrm{Tr} \Tilde{e}^I \circ \Tilde{e}^J =
            \eta^{IJ} \\
  \left( e_I,e_J\right) &= &\mathop\mathrm{Tr} e_I \circ e_J = \eta_{IJ} \\
  (e_I,\Tilde{e}^J) &= &Tr e_I \circ \Tilde{e}^J = \delta_I^J 
\end{eqnarray}
where $\eta_{IJ} $ is diagonal and equal to $\delta_{IJ}$ for
Euclidean (compact) elements and equal to $-\delta_{IJ}$ for non
Euclidean (noncompact) elements and will be explicitly given
below. Furthermore we will label the basis elements of the Jordan
algebra $J$ such that $e_1, e_2$ and $e_3$ are the three irreducible
idempotents of $J$ and the identity element $\mathbb{I}$ is simply
\begin{equation}
  \mathbb{I} = e_1 + e_2 + e_3
\end{equation}

The action of the generators of quasiconformal group $QConf(J)$ on the
space $\mathcal{T} = \mathcal{F}(J) \oplus \mathbb{R}$ with
coordinates $q_0$, $q_I$, $p^0$, $p^I$ of $\mathcal{F}(J)$ plus an extra
singlet coordinate $x \in \mathbb{R}$, twisted by a unitary character
$\nu$, is given by the following differential operators:
\begin{eqnarray} \label{generators}
K&=&\partial_x \\
U_0 &=& \partial_{p^0} + q_0 \partial_x  \\
U_I &=& -  \partial_{p^I} +  q_I \partial_x \\
V^0 &=& \partial_{q_0} - p^0 \partial_x \\
V^I &=&  \partial_{q_I} +  p^I \partial_x \\
%\end{eqnarray}
%\begin{eqnarray}
   R_I &= & -\sqrt{2} C_{IJK} p^K \partial_{q_K} - \sqrt{\frac{3}{2}}
   \left( p^0 \partial_{p^I} + q_I \partial_{q_0}\right) \\
   \Tilde{R}^I & =& \sqrt{2} C^{IJK} q_J \partial_{p^K} +
   \sqrt{\frac{3}{2}} \left( q_0 \partial_{q_I} + p^I \partial_{p^0}
   \right) \\
   {R_I}^J& = &\frac{3}{2} {\delta_I}^J \left( p^0 \partial_{p^0} -
             q_0 \partial_{q_0}\right) + \\ \nonumber && \frac{3}{2}
             \left( {\delta_I}^N {\delta_K}^J - \frac{4}{3} C_{IKL}
             C^{JNL} \right) \left( q_N \partial_{q_K} - p^K
             \partial_{p^N}\right) \\
  \mathcal{R} &=& \frac{1}{n_V} {R_I}^I = \frac{3}{2} \left( p^0
                       \partial_{p^0} - q_0 \partial_{q_0}\right) +
                       \frac{1}{2} \left( p^I \partial_{p^I} - q_I
                       \partial_{q^I}\right) \\ && \nonumber \\
   \Delta & = &=-\mathcal{D}= -\left( p^0 \partial_{p^0} + p^I
      \partial_{p^I} + q_0 \partial_{q_0} + q_I \partial_{q_I} - \nu
     \right)-2x\partial_{x} \\
  \nonumber && \\
  \Tilde{K} &= & x \left( p^0 \partial_{p^0} + p^I \partial_{p^I} +
               q_0 \partial_{q_0} + q_I \partial_{q_I} - \nu \right) +
               \left( x^2 + I_4 \right) \partial_x \nonumber \\ & & +
               \frac{1}{2} \left( \frac{ \partial{I_4}}{\partial p^0}
               \partial_{q_0} - \frac{ \partial{I_4}}{\partial q_0}
               \partial_{p^0} + \frac{ \partial{I_4}}{\partial q_I}
               \partial_{p^I} - \frac{ \partial{I_4}}{\partial p^I}
               \partial_{q_I} \right) \label{generators_last}
\end{eqnarray}

The vertical five grading is determined  by the adjoint action of $\mathcal{D}$
\begin{eqnarray}
\left[ \mathcal{D}, \left(\begin{array}{c} U_0 \\U_I \\V^I  \\ V^0 \end{array}\right) \right] =
 \left(\begin{array}{c} U_0 \\U_I \\V^I \\ V^0 \end{array}\right)
\end{eqnarray}
The vertical grade $-1$ generators are obtained from the grade +1
generators by commutation with the grade $-2$ generator $\Tilde{K}$
\begin{eqnarray}
    \Tilde{U}_0 &=& \left[ U_0, \Tilde{K}\right] \\
    \Tilde{V}^0 &=& \left[ V^0, \Tilde{K} \right] \\
    \Tilde{U}_I &=& \left[ U_I, \Tilde{K}\right] \\
    \Tilde{V}^I &=& \left[ V^I, \Tilde{K} \right]
\end{eqnarray}
and satisfy
\begin{eqnarray}
[ \mathcal{D}, \left(\begin{array}{c} \Tilde{U}_0 \\ \Tilde{U}_I\\
\Tilde{V}^I \\ \Tilde{V}^0 \end{array}\right) ] =
-\left(\begin{array}{c} \Tilde{U}_0 \\ \Tilde{U}_I \\ \Tilde{V}^I \\
\Tilde{V}^0 \end{array}\right)
\end{eqnarray}
The remaining non-vanishing commutation relations of Lie algebra of $QConf(J)$  are as follows:
\begin{eqnarray}
 \left[ K , \Tilde{K} \right] &=& \Delta \\
 \left[ \Delta, K \right] &=& -  2K \\
 \left[ \Delta, \Tilde{K} \right] &=& 2 \Tilde{K} \\
\left[ U_I,V^J \right] &=& -2 \delta_I^J K \\
\left[ U_0, V^0 \right] &=& -2 K \\
\left[ K, \Tilde{U}_0 \right] &=& U_0 \\
\left[ K, \Tilde{U}_I \right] &=& U_I \\
\left[ K, \Tilde{V}^I \right] &=& V^I \\
\left[ K, \Tilde{V}^0 \right] &=& V^0\\
\left[ \Tilde{U}_I , \Tilde{V}^J \right] &=&- 2 \delta_I^J \Tilde{K} \\
\left[ \Tilde{U}_0 , \Tilde{V}^0 \right] &=&- 2  \Tilde{K} \\
%\begin{equation}
    \left[ U_0 ,\Tilde{V}^0 \right] &=& -2 \mathcal{R} + \mathcal{D} \\
%\end{equation}
%\begin{equation}
   \left[ V^0 ,\Tilde{U}_0 \right] &=& -2 \mathcal{R} - \mathcal{D} 
%\end{equation}
\end{eqnarray}
\begin{eqnarray}
\left[ R_I^J, R_K \right] &=&
     \frac{3}{2} \Lambda_{IK}^{JL} R_L \\
\left[ R_I^J, \tilde{R}^L \right]& =&
    -\frac{3}{2} \Lambda_{IL}^{JK} \tilde{R}^L \\
\left[ R_I^J, U_K \right] &=&
     \frac{3}{2} \Lambda_{IK}^{JL} U_L -
     \frac{3}{2} \delta_I^J U_K \\
\left[ R_I^J, V^K \right]& =&
     -\frac{3}{2} \Lambda_{IL}^{JK} V^L +
     \frac{3}{2} \delta_I^J V^K \\
\left[ R_I^J, \tilde{U}_K \right] &=&
      \frac{3}{2} \Lambda_{IK}^{JL} \tilde{U}_L -
      \frac{3}{2} \delta_I^J \tilde{U}_K \\
\left[ R_I^J, \tilde{V}^K \right] & =&
     -\frac{3}{2} \Lambda_{IL}^{JK} \tilde{V}^L +
     \frac{3}{2} \delta_I^J \tilde{V}^K \\
\left[ R_I, \tilde{R}^J \right] &=& - R_I^J \\
%\end{eqnarray}
%\begin{eqnarray}
\left[ U_0, \tilde{V}^I \right] &= &
      2\sqrt{\frac{2}{3}} \tilde{R}^I \\
\left[ \tilde{U}_0, V^I \right] &=&
      -2\sqrt{\frac{2}{3}} \tilde{R}^I  \\
\left[ V^0, \tilde{U}_I \right] &=&
      -2\sqrt{\frac{2}{3}} R_I    \\
\left[ \tilde{V}^0, U_I \right] &=&
      -2\sqrt{\frac{2}{3}} R_I             \\
\left[ U_I, \tilde{V}^J \right] &=&
       \frac{4}{3} R_I^J -
       \delta_I^J \left( \Delta + 2 {\mathcal R} \right) \\
%\end{eqnarray}
%\begin{eqnarray}
\left[ \tilde{U}_I, V^J \right] &=&
       -\frac{4}{3} R_I^J -
       \delta_I^J \left( \Delta - 2 {\mathcal R} \right)  \\
\left[ U_I, \tilde{U}_J \right] &=&
     - \frac{4}{3} \sqrt{2} C_{IJK} \tilde{R}^K \\
\left[ V^I, \tilde{V}^J \right] &=&
     - \frac{4}{3} \sqrt{2} C^{IJK} R_K   \\
\left[ V^I, R_J \right] &=&
      -\sqrt{\frac{3}{2}} \delta^I_J V^0   \\
\left[ \tilde{V}^I, R_J \right] &=&
      -\sqrt{\frac{3}{2}} \delta^I_J \tilde{V}^0  \\
\left[ \tilde{U}_I, \tilde{R}^J \right] &=&
      -\sqrt{\frac{3}{2}} \delta_I^J \tilde{U}_0  \\
\left[ U_I, \tilde{R}^J \right] &=&
      -\sqrt{\frac{3}{2}} \delta_I^J  U_0 \\
\end{eqnarray}
where
\begin{equation}
   \Lambda_{KL}^{IJ} := \delta_K^I \delta_L^J + \delta_L^I \delta^J_K - \frac{4}{3} C^{IJM} C_{KLM}
\end{equation}

There is a distinguished $SL(3,\mathbb{R})$ subgroup of $QConf(J)$
whose centralizer is the Lorentz group $Str_0(J)$ generated by $L_I^J$, 
whose generators are $K$, $\Tilde{K}$, $U_0$, $\Tilde{U}_0$, $V^0$,
$\Tilde{V}^0$, $\mathcal{R}$ and $\mathcal{D}$ .  Its maximal compact
subgroup $SO(3) \subset SL(3,\mathbb{R})$ is generated by
\begin{eqnarray}
T_1 &:= & \frac{1}{\sqrt{2}} \left( U_0 - \Tilde{V}^0 \right)  \\
T_2 & := & \frac{1}{\sqrt{2}} \left( V^0 + \Tilde{U}_0 \right) \\
T_3 & := & - \left( K + \Tilde{K} \right)
\end{eqnarray}
that satisfy 
\begin{equation}
\left[ T_i , T_j \right] =\epsilon_{ijk} T_k
\end{equation}
where $i,j,k=1,2,3$.
The generators of the maximal compact subgroup $K$ of $QConf(J)$ are
\begin{eqnarray}
&(U_I-\eta_{IJ} \Tilde{V}^J), \nonumber \\ \nonumber
&(V^I + \eta^{IJ} \Tilde{U}_J), \\ \nonumber & (R_I + \eta_{IJ}\Tilde{R}^J ), \\ \nonumber & (R_I^J-\eta_{IK} \eta^{JL}R_L^K), \\
& (\Tilde{U}_0 + V^0 ), \\ \nonumber
& (-U_0 + \Tilde{V}^0), \\ \nonumber
 & (K + \Tilde{K})
\end{eqnarray}
where $I,J,...= 1,2,..., n_V$ where $n_V$ is the dimension of $J$
\footnote{For the supergravity theories whose C-tensors are determined
by a Jordan algebra, $n_V$ is the number of vector fields in 5
dimensions. hence the notation.}.

\subsection{Quadratic Casimir Operators of Quasiconformal Lie algebras}
\renewcommand{\theequation}{\arabic{section}.{\arabic{subsection}}.\arabic{equation}}
\setcounter{equation}{0}
The generators $L_I^J$ of the reduced structure (Lorentz) group of a
Jordan algebra are given by the traceless components of $R_I^J$:
\begin{equation}
L_I^J = R_I^J - \frac{1}{n_V} \delta_I^J ( R_K^K) = R_I^J - \delta_I^J \mathcal{R}
\end{equation}
The quadratic Casimir operator of the quasiconformal group $QConf(J)$
of a simple Jordan algebra $J$ of degree three can then be written in
a general form involving a single parameter $\alpha$, valid both for
Euclidean and non-Euclidean $J$:

\begin{eqnarray}
\mathcal{C}_2 &= &\alpha L_I^J L_J^I - \frac{4}{3} ( \mathcal{R}^2 +
\Tilde{R}^I R_I + R_I \Tilde{R}^I ) + ( U_0 \Tilde{V}^0 + U_I
\Tilde{V}^I + \Tilde{V}^0 U_0 + \Tilde{V}^I U_I ) \\ \nonumber && -
(\Tilde{U}_0 V^0 + \Tilde{U}_I V^I + V^0 \Tilde{U}_0 + V^I \Tilde{U}_I
) - 2 ( K \Tilde{K} + \Tilde{K} K) + \Delta^2
\end{eqnarray}
where $\alpha$ takes on the following  values for different quasiconformal Lie groups $QCG(J)$:
\begin{eqnarray}
\alpha(F_{4(4)})= \frac{16}{45} \\ \nonumber
\alpha(E_{6(6)})= \frac{8}{27} \\ \nonumber
\alpha(E_{7(7)})= \frac{2}{9} \\ \nonumber
\alpha(E_{8(8)})= \frac{4}{27} \\ \nonumber
%\alpha(SO(n_V+2,4)) = \frac{4}{9} 
\end{eqnarray}
As for the quasiconformal groups $QConf(\mathbb{R} \oplus
\Gamma_{(n,m)})= SO(n+3,m+3)$ associated with the generic family of
reducible Jordan algebras $J= \mathbb{R} \oplus \Gamma_{(n,m)}$ the
quadratic Casimir can be written in the form:
\begin{eqnarray}
\mathcal{C}_2 ( SO(n+3,m+3))  &= &\frac{4}{9} R_I^J R_J^I - 
   \frac{4}{3} ( \Tilde{R}^I R_I + R_I \Tilde{R}^I ) - 
   \frac{(n+m-2)}{9}  (R_2^2 +R_3^3)^2 \nonumber \\ &&
   + ( U_0 \Tilde{V}^0 + U_I \Tilde{V}^I + \Tilde{V}^0 U_0 + 
   \Tilde{V}^I U_I ) \\ \nonumber &&  - (\Tilde{U}_0 V^0 + 
   \Tilde{U}_I V^I + V^0 \Tilde{U}_0 + V^I \Tilde{U}_I )
   - 2 ( K \Tilde{K} + \Tilde{K} K) + \Delta^2 
\end{eqnarray}
The dimension of $J=\mathbb{R} \oplus \Gamma_{(n,m)}$ is
\begin{equation}
n_V=n+m+1
\end{equation}

The quadratic Casimir operators for all the quasiconformal groups
$QConf(J)$ reduce to a c-number whose value can be expressed
universally in terms of the twisting parameter and the dimension $n_V$
of the Jordan algebra $J$ as :
\begin{equation}
\mathcal{C}_2 ( QCG(J)) = \nu (\nu + 2n_V + 4 ) 
\end{equation}
As a consequence the representations induced by the quasiconformal
action of $QConf(J)$ on the space of square integrable functions
$L^2(p^I,q_I,x) $ of $(n_V+1)$ coordinates $q_I$ , $(n_V+1)$ momenta
and the singlet coordinate $x$ are unitary representations belonging
to the principle series under the scalar product
\begin{equation}
\langle f|g\rangle = \int \bar{f}(p,q,x) g(p,q,x) dp dq dx 
\end{equation}
for 
\begin{equation}
\nu = -(n_V+2) + i \rho 
\end{equation}
where $\rho \in \mathbb{R}$. For special discrete values of the
twisting parameter $\nu$ one obtains representations belonging to the
discrete series and their continuations. Typically these
representations arise as submodules of the Verma modules obtained by
the action of noncompact generators on the spherical vectors of
quasiconformal group actions. For the quasiconformal groups of
Euclidean Jordan algebras these special representations include the
quaternionic discrete representations and their continuations
\cite{Gunaydin:2007qq,mgop}, which can be realized over the space of
holomorphic functions of complexified quasiconformal
coordinates. These holomorphic coordinates can be identified with the
natural complex coordinates of the twistor spaces associated with the
quaternionic symmetric spaces
\begin{equation*}
 \frac{ QConf(J)}{ \widetilde{Conf}(J) \times SU(2)} 
\end{equation*}
when $J$ is Euclidean.  For rank two quaternionic quasiconformal
groups $SU(2,1)$ and $G_{2(2)}$ these representations were studied in
\cite{Gunaydin:2007qq}.

\section{Spherical Vectors  of Quasiconformal Groups associated with Split Non-Euclidean  Jordan algebras of Degree Three}
\renewcommand{\theequation}{\arabic{section}.{\arabic{subsection}}.\arabic{equation}}
\setcounter{equation}{0}
A Jordan algebra $J$ of degree three admits three mutually orthogonal
irreducible idempotents $\mathbb{P}_1$,$\mathbb{P}_2$,$\mathbb{P}_3$:
\begin{equation}
  \mathbb{P}_i \circ \mathbb{P}_j = \delta_{ij} \mathbb{P}_i
\end{equation}
\begin{equation}
  Tr(\mathbb{P}_i) = 1 \hspace{1cm} i,j,..=1,2,3 \nonumber
\end{equation}
As stated earlier in our labelling of the basis elements of split
Jordan algebras of degree three we identify $e_i$ with $\mathbb{P}_i$
for $i=1,2,3$.  By the action of the automorphism group $Aut(J)$ one
can ``diagonalize'' a general element $\mathbf{x}\in J$
\begin{equation}
  Aut(J): \;\;\; \mathbf{x} \longrightarrow \mathbf{x}_D= \lambda_1 e_1 + \lambda_2 e_2 + \lambda_3 e_3
\end{equation}
The cubic norm of $\mathbf{x}$ is therefore
\begin{equation}
  \mathcal{N}(\mathbf{x}) = 3 \sqrt{3} (\lambda_1 \lambda_2 \lambda_3 )
\end{equation}

This holds true for Euclidean as well as for split non-Euclidean
Jordan algebras of degree three (assuming $n_V \geqslant 3$).  The
quasiconformal group of the Jordan subalgebra generated by the
irreducible idempotents is $SO(4,4)$ which is the U-duality group of
the STU model and was studied in \cite{Gunaydin:2009dq}. This Jordan
subalgebra is isomorphic to the Euclidean algebra $(\mathbb{R} \oplus
\Gamma_{(1,1)})$, which is a subalgebra of all the generic family of
Jordan algebras $(\mathbb{R}\oplus \Gamma_{(m,n)})$ for $m,n >1$.  The
simple split non-Euclidean Jordan algebras all have $J_3^{\mathbb{R}}$
as a subalgebra and hence its quasiconformal group $F_{4(4)}$ is a
subgroup of the split exceptional quasiconformal groups $E_{6(6)}$,
$E_{7(7)}$ and $E_{8(8)}$ of the split simple Jordan algebras
$J_3^{\mathbb{C}_s}$, $J_3^{\mathbb{H}_s}$ $J_3^{\mathbb{O}_s}$. Now
the quasiconformal groups of all Euclidean Jordan algebras of degree
three are of the quaternionic real form , in particular the groups
$SO(4,4)$ and $F_{4(4)}$. 

Unitary representations induced by the action of quaternionic
quasiconformal groups with unitary character $\nu$ include the
quaternionic discrete series representations of Gross and Wallach
\cite{MR1421947} as was shown explicitly for rank two cases in
\cite{Gunaydin:2007qq}.  The explicit expressions for the spherical
vectors of quasiconformal realizations of $SU(2,1)$ and $G_{2(2)}$
were essential to establish this result \cite{Gunaydin:2007qq}.  The
quaternionic discrete series representations and their continuations
appear as submodules in the Verma modules generated by the action of
noncompact generators on the spherical vectors for special discrete
values of the parameter $\nu$ that is the twisting parameter in the
quasiconformal group action.  The spherical vectors for all
quaternionic quasiconformal groups defined by Euclidean Jordan
algebras of degree three were given in \cite{Gunaydin:2009dq}. In this
section we shall extend these results to quasiconformal groups of all
split non Euclidean Jordan algebras of degree three. The study of
discrete series representations induced by the corresponding
quasiconformal group actions will be subjects of separate studies.

Now the spherical vector of quasiconformal group action of $QConf(J)$
twisted by a unitary character $\nu$ is a function 
$\Phi_{\nu}(p,q,x)$ of $2n_v+3$ variables $q_0$,$q_I$,$p^0$,$p^I$ and $x$
that is annihilated by all the generators $\mathfrak{K}_M$ of the
maximal compact subgroup $K$ of $QConf(J)$:
\begin{equation}
\mathfrak{K}_M \Phi_{\nu}(p,q,x)=0
\end{equation}
In \cite{Gunaydin:2009dq} we presented the spherical vectors of all
quasiconformal groups associated with Euclidean Jordan algebras of
degree three.  The spherical vector of a general quasiconformal group
$QConf(J)$ associated with a split Non-Euclidean Jordan algebra $J$
can be written in the form:
\begin{eqnarray} \label{sphericalvector}
    \Phi_{\nu}(p,q,x) &=& [ (1+x^2+ I^S_2 -I_4)^2 - (I^S_2)^2  +8 I_4
      + \frac{1}{2} I^S_6 + 8 x J^S_4 + \frac{4}{81} H^S_4 ]^{\frac{\nu}{4}}
\end{eqnarray}
where
\begin{equation}
   I^S_2 = (p^0)^2 + (q_0)^2 + p^I \eta_{IJ} p^J + q_I \eta^{IJ} q_J
\end{equation}
\begin{eqnarray}
I_4 &=&
    \left(p^0 q_0 -  p^I q_I \right)^2 - \frac{4}{3}  C_{IJK} p^J p^K C^{ILM} q_L q_M \\  \nonumber && +
             \frac{4}{3\sqrt{3}} p^0 C^{IJK} q_I q_J q_K + \frac{4}{3\sqrt{3}} q_0 C_{IJK} p^I p^J p^K
\end{eqnarray}
\begin{equation}
      J^S_4 = \frac{1}{4} \left(p^0 \frac{\partial{I_4}}{\partial{q_0}} -
                              q_0 \frac{\partial{I_4}}{\partial{p^0}}
			      +q_I \eta^{IJ} \frac{\partial{I_4}}{\partial{p^J}} -
			      p^I \eta_{IJ}\frac{\partial{I_4}}{\partial{q_J}}
                        \right)
\end{equation}
\begin{equation}
   I^S_6 = \left( \frac{\partial{I_4}}{\partial{p^0}}\right)^2 +\left(
       \frac{\partial{I_4}}{\partial{q_0}}\right)^2 +\left(
       \frac{\partial{I_4}}{\partial{p^I}}\right) \eta^{IJ}\left(
       \frac{\partial{I_4}}{\partial{p^J}}\right) + \left(
       \frac{\partial{I_4}}{\partial{q_I}}\right)\eta_{IJ} \left(
       \frac{\partial{I_4}}{\partial{q_J}}\right) + 4I_4 I^S_2
\end{equation}
\begin{eqnarray}
    H^S_4 & = &27 \eta^{IJ} \left( (p^{\#})_I -\sqrt{3} p^0 q_I \right) \left( (p^{\#})_J -\sqrt{3} p^0 q_J \right) \nonumber \\ &&  + 27 \eta_{IJ} \left( (q^{\#})^I -\sqrt{3} q_0 p^I \right) \left( (q^{\#})^J -\sqrt{3} q_0 p^J \right)  \\ &&
    + 54\left((q^{\#})^I -\sqrt{3} q_0 p^I \right) \left( (p^{\#})_I -\sqrt{3} p^0 q_I \right) \nonumber  + C^S_4(J)
\end{eqnarray}
where $(p^{\#})_I= C_{IJK}p^Jp^K$ and $(q^{\#})^I= C^{IJK}q_Jq_K$.
$C^S_4(J)$ is the ``correction'' term that vanishes when restricted
to the subalgebra $SO(4,4)$ and has a different form for simple
Jordan algebras and non-simple ones.  For simple split non-Euclidean
Jordan algebras $J$ of degree three the quartic correction term
$C^S_4(J)$ is given by
\begin{eqnarray}
      C^S_4 (J_3^{\mathbb{A}_S}) & = & 81 \left( \mathop{Tr}\left[ M_0(p)\circ \Tilde{M}_0(q)\right]\right)^2 +
                \frac{81}{2} \mathop{Tr} \left[ M_0(p)^2 \right] \mathop{Tr} \left[ \Tilde{M}_0(q)^2 \right] \\ \nonumber &&
		-243 \mathop{Tr} \left[\{M_0(p), \Tilde{M}_0(q), M_0(p)\}\circ \Tilde{M}_0(q) \right]
\end{eqnarray}
where
\begin{eqnarray}
 M_0(q)  &=& M(q) - \frac{1}{3} Tr M(q) \\   M(q) & = & e^I q_I \in J_3^{\mathbb{A}_S}
 \end{eqnarray}
and $\Tilde{M}(q) = M^*(q)$ where $*$ is the conjugation that replaces
the "imaginary" units $j_\mu^S$ that square to $+1$ by $-j_\mu^S$ in
the underlying split composition algebra $\mathbb{A}^S$.  $\{A,B,C\}$
denotes the Jordan triple product
\begin{equation}
\{A,B,C\} = A\circ ( B \circ C ) + C \circ ( B \circ A ) - (A\circ C) \circ B
\end{equation}
For special Jordan algebras with the Jordan product
\begin{equation*}
   A \circ B = \frac{1}{2} (A B + B A )
\end{equation*}
one finds
\begin{equation}
   \left\{ A,B,A \right\} = A B A
\end{equation}
Therefore for Jordan algebras $J_3^{\mathbb{C}_S}$ and
$J_3^{\mathbb{H}_S}$ the term $C^S_4(J)$ can be written as
\begin{eqnarray}
          C^S_4 (J_3^{\mathbb{A}_S}) &= &
             81 \left(\mathop{Tr} \left[ M_0(p) \Tilde{M}_0(q) \right] \right)^2 +
             \frac{81}{2} \mathop{Tr} \left[ M_0(p)^2 \right] \mathop{Tr} \left[ \Tilde{M}_0(q)^2 \right] \\ \nonumber
             && -243 \mathop{Tr} \left[M_0(p) \Tilde{M}_0(q) M_0(p) \Tilde{M}_0(q) \right] \nonumber
\end{eqnarray}

For the generic nonsimple Jordan algebras $(\mathbb{R} \oplus
\Gamma_{(n,m)})$ ($ m,n \geqslant 1$ of degree three the cubic form is
\begin{equation}
\mathcal{N}(q) =C^{IJK} q_Iq_Jq_K=  \frac{3\sqrt{3}}{2} q_1 [ 2q_2q_3 - \eta^{\hat{I}\hat{J}} q_{\hat{I}} q_{\hat{J}}]
\end{equation}
and the quartic correction term $C^S_4$ that appears in  $H^S_4$ is given by
\begin{eqnarray}
        C^S_4 ( \mathbb{R}\oplus \Gamma_{(n,m)})  &= &
      - \frac{81}{2} \left\{ (p^2 - p^3)(q_2 - q_3) + 2 p^{\hat{I}} q_{\hat{I}} \right\}^2  \\ &&
    + \frac{81}{2} \left\{ (p^2 -p^3)^2 + 2 \eta_{\hat{I}\hat{J}} p^{\hat{I}} p^{\hat{J}} \right\}
     \left\{ (q_2 - q_3)^2  + 2 \eta^{\hat{I}\hat{J}} q_{\hat{I}} q_{\hat{J}} \right\} \nonumber
\end{eqnarray}
where $\hat{I},\hat{J},...=4,5,..., (m+n+1) $.

We should perhaps recall that the Euclidean Jordan algebra
$J_3^{\mathbb{R}} $ is a subalgebra of all simple split Jordan
algebras of degree three. An element of $J_3^{\mathbb{R}} $ can be
written as
\begin{equation}
M (p)  = \frac{1}{\sqrt{2}}\left( {\begin{array}{*{20}c}
   \sqrt{2} p^1 & p^6 & p^5  \\
   p^6 & \sqrt{2}p^2 & p^4  \\
   p^5 & p^4 &\sqrt{2}p^3  \\
\end{array}} \right)
\end{equation}
and its cubic norm is simply
\begin{eqnarray}
       \mathcal{N}(M(p)) &=& 3 \sqrt{3} Det M= C_{IJK} p^Ip^Jp^K \\ \nonumber && = 3 \sqrt{3} \left\{ p^1 p^2 p^3 -
           \frac{1}{2} \left[  p^1 (p^4)^2 + p^2 (p^5)^2 + p^3 (p^6)^2 \right]
         + \frac{1}{\sqrt{2}} p^4 p^5 p^6 \right\}
\end{eqnarray}

\subsection{Quasiconformal Group $E_{8(8)}$ }

The quasiconformal group associated with the split exceptional Jordan
algebra $J_3^{\mathbb{O}_S}$ is the split exceptional group
$E_{8(8)}$. A general element of $J_3^{\mathbb{O}_S}$ can be written
as
\begin{equation}
M (p)  = \frac{1}{\sqrt{2}}\left( {\begin{array}{*{20}c}
   \sqrt{2} p^1 & P_S^6 & \bar{P}_S^5  \\
   \bar{P}_S^6 & \sqrt{2}p^2 & P^4  \\
   P_S^5 & \bar{P}_S^4 &\sqrt{2}p^3  \\
\end{array}} \right)
\end{equation}
where $P_S^4$, $P_S^5$ and $P_S^6$ are split octonions. The cubic norm of
$M(p)$ is normalized to be
\begin{equation}
    \mathcal{N}(M(p)) = 3 \sqrt{3}\{ p^1 p^2 p^3 -
        \frac{1}{2} \left( p^1 P_S^4 \bar{P}_S^4 + p^2 P_S^5 \bar{P}_S^5 + p^3 P_S^6 \bar{P}_S^6 \right)
        + \frac{1}{\sqrt{2}} Re (P_S^4 P_S^5 P_S^6) \}
\end{equation}
where $Re(X_S)$ denotes the real part of $X _S\in \mathbb{O}_S$ and
$\bar{X}_S$ is the octonion conjugate of $X_S$.  If we expand the
elements $P_S^4$, $P_S^5$ and $P_S^6$ in terms of their real components
\begin{eqnarray}
 P_S^4 = p^4 + p^{4+3i} j_i +  p^{4+3\mu} j_\mu^s  \nonumber \\
 \bar{P}_S^4 = p^4 - p^{4+3i} j_i -  p^{4+3\mu} j_\mu^s \nonumber \\
 P_S^5 = p^5 + p^{5+3i} j_i +p^{5+3\mu} j_\mu^s\nonumber \\
 \bar{P}_S^5 = p^5 -p^{5+3i} j_i -p^{5+3\mu} j_\mu^s \\
 P_S^6 = p^6 + p^{6+3i} j_i + p^{6+3\mu} j_\mu^s\nonumber \\
 \bar{P}_S^6 = p^6 - p^{6+3i} j_i - p^{6+3\mu} j_\mu^s\nonumber
\end{eqnarray}
where the indices $i$ and $\mu$ are summed over with $i$ running over
$1,2,3$ and $\mu$ running over $4,5,6,7$,
we can write the cubic norm as
\begin{eqnarray} \label{cubicform_E8}
      \mathcal{N}(M(p))& = &3 \sqrt{3} \Bigg\{ p^1 p^2 p^3 -  \frac{1}{2} p^1 [( p^4)^2  + p^{4+3i} p^{4+3i} - p^{4+3\mu} p^{4+3\mu}] \nonumber \\ &&
       -\frac{1}{2} p^2 [ (p^5)^2 + p^{5+3i} p^{5+3i} - p^{5+3\mu} p^{5+3\mu} ] -\frac{1}{2} p^3 [ (p^6)^2 + p^{6+3i} p^{6+3i}- p^{6+3\mu} p^{6+3\mu} ] \nonumber  \\ &&
      + \frac{1}{\sqrt{2}} \Big\{ p^4 p^5 p^6   - p^4 \left( p^{(5+3i)} p^{(6+3i)}- p^{(5+3\mu)} p^{(6+3\mu)} \right) - p^5 \left( p^{(4+3i)}  p^{(6+3i)} - p^{(4+3\mu)} p^{(6+3\mu)} \right)  \nonumber  \\ &&   - p^6 \left(p^{(4+3i)}  p^{(5+3i)} -p^{(4+3\mu)}  p^{(5+3\mu)}\right)\Big\} 
      -\frac{1}{\sqrt{2}} \epsilon_{ijk} p^{4+3i} p^{5+3j} p^{6+3k}  \\ \nonumber &&
       +\frac{1}{\sqrt{2}} \eta_{i\mu\nu} p^{4+3i} p^{5+3\mu} p^{6+3\nu} 
        +\frac{1}{\sqrt{2}} \eta_{\mu i \nu} p^{4+3\mu} p^{5+3i} p^{6+3\nu} 
         +\frac{1}{\sqrt{2}} \eta_{\mu\nu i} p^{4+3\mu} p^{5+3\nu} p^{6+3i}       
      \Bigg\} \nonumber
\end{eqnarray}
With the above labelling of the basis elements we have 
\eqn
\eta_{AB}& =& \delta_{AB}  \\ \nonumber
\eta_{RS}& =& - \delta_{RS} \\
\eta_{AR}& =& 0 \nonumber
\enn 
where $A,B,..= 1,2, ..., 15$ and $ R,S,..= 16,17,...,27 $.
The conjugate elements $\Tilde{M}(p)$ of $J_3^{\mathbb{O}_S}$ obtained by conjugation $*$ is given by
\begin{equation}
M (p)  = \frac{1}{\sqrt{2}}\left( {\begin{array}{*{20}c}
   \sqrt{2} p^1 &( P_S^6)^* & (\bar{P}_S^5)^*  \\
   (\bar{P}_S^6)^* & \sqrt{2}p^2 & (P^4)^*  \\
  ( P_S^5)^* & (\bar{P}_S^4)^* &\sqrt{2}p^3  \\
\end{array}} \right)
\end{equation}
where $*$ is the conjugation under which
\begin{equation*} (j_i)^* = j_i \end{equation*}
\begin{equation*}( j_\mu^s)^* = - j_\mu^s \end{equation*}
where $i=1,2,3$ and $\mu=4,5,6,7$. 
Thus we find 
\begin{equation*}
Tr M(p) \circ \Tilde{M} (p) = \sum_{I=1}^{27} p^Ip^I 
\end{equation*}

The explicit expressions for the generators of quasiconformal group
$E_{8(8)}$ and its spherical vector are obtained by substituting the
expressions for the cubic form \ref{cubicform_E8} and the metric
$\eta_{IJ}$ in (\ref{generators}-\ref{generators_last})  and in \ref{sphericalvector}. The generators of the maximal compact subgroup $SO(16)$
of $E_{8(8)}$ are
\begin{eqnarray}
 \label{maximalcompact_E8}
(U_A - \Tilde{V}^A ) \nonumber \\ \nonumber
(U_R +\Tilde{V}^R ) \\ \nonumber
(V^A + \Tilde{U}_A ) \\
(V^R - \Tilde{U}_R ) \\ \nonumber
(R_A + \Tilde{R}^A ) \\ \nonumber
(R_R - \Tilde{R}^R ) \\ \nonumber
(R_A^B-R_B^A ) \\ \nonumber
(R_R^S - R_S^R ) \\ \nonumber
(R_A^S + R_S^A ) \\ \nonumber
( \Tilde{U}_0 + V^0) \\ \nonumber
(-U_0 + \Tilde{V}^0 ) \\ \nonumber
(K + \Tilde{K} )  \nonumber
\end{eqnarray}

$SU(8)$ subgroup of $SO(16)$ generated by $(R_A^B-R_B^A)$, $(R_R^S -
R_S^R)$, $(R_A^S + R_S^A)$, $(R_A + \Tilde{R}^A)$ and $(R_R - \Tilde{R}^R )$ 
act linearly in the quasiconformal action of $E_{8(8)}$
on the 57 dimensional space with coordinates $p^0$,$p^I$,$q_0$,$q_I$ and
$x$. Thus we shall refer to the coset space
\begin{equation}
\mathcal{K}(QConf(J_3^{\mathbb{O}_S}) = \frac{SO(16)}{SU(8)} 
\end{equation}
as the ``quasiconformal compactification'' of this 57 dimensional
space \cite{mg2009}.  As explained in detail in \cite{Gunaydin:2000xr}
one can embed the subgroup $E_{7(7)}$ inside $E_{8(8)}$ in essentially
three different ways. The $E_{7(7)}$ subgroup generated by the grade
zero generators (with respect to $\mathcal{D}$) acts linearly. In
addition we have two different subgroups, which we label as
$E^q_{7(7)}$ and $E_{7(7)}^p $ that act as the nonlinear conformal
group of $J_3^{\mathbb{O}_S}$ on the coordinates and momenta ,
respectively. The conformal compactification of the corresponding 27
dimensional space is \cite{mg2009}
\begin{equation}
\mathcal{K}(Conf(J_3^{\mathbb{O}_S}) = \frac{SU(8)}{USp(8)} 
\end{equation}

It is interesting to compare $E_{8(8)}$ as the quasiconformal group
associated with the split exceptional Jordan algebra with $E_{8(-24)}$
as the quasiconformal group of the Euclidea exceptional Jordan algebra
defined over the division algebra of real octonions.  $E_{8(-24)}$ has
the maximal compact subgroup $E_7 \times SU(2)$. The quasiconformal
action $E_{8(-24)}$ extends naturally to a holomorphic quasiconformal
action on the complex coordinates of the corresponding twistor space
\cite{Gunaydin:2007qq}
\begin{equation*} 
  \frac{E_{8(-24)} \times SU(2)} { E_7 \times SU(2) \times U(1)} = \frac{E_{8(-24)}} { E_7 \times U(1)}
\end{equation*}
The quasiconformal compactification of the 57 dimensional space on
which $E_{8(-24)}$ acts is the coset space\cite{mg2009}
\begin{equation}
\mathcal{K}(QConf(J_3^{\mathbb{O}}) = \frac{E_7 \times SU(2)}{E_6 \times U(1)}
\end{equation}
The subgroup $E_{7(-25)}$ has similarly three inequivalent embeddings
inside $E_{8(-24)}$. The conformal compactification of the 27
dimensional space on which $E_{7(-25)}= Conf(J_3^{\mathbb{O}})$ acts
nonlinearly is the coset space
\begin{equation}
\mathcal{K}(Conf(J_3^{\mathbb{O}}) = \frac{E_6}{F_4}\times S^1 
\end{equation}

\subsection{Quasiconformal Group $E_{7(7)}$ } 

Quasiconformal group associated with the Jordan algebra
$J_3^{\mathbb{H}_S} $ over the split quaternions $\mathbb{H}_S$ is
$E_{7(7)}$. A general element of $J_3^{\mathbb{H}_S} $ can be written
in the form
\begin{equation}
M (p)  = \frac{1}{\sqrt{2}}\left( {\begin{array}{*{20}c}
   \sqrt{2} p^1 & P_S^6 & \bar{P}_S^5  \\
   \bar{P}_S^6 & \sqrt{2}p^2 & P^4  \\
   P_S^5 & \bar{P}_S^4 &\sqrt{2}p^3  \\
\end{array}} \right)
\end{equation}
where $P_S^4$, $P_S^5$ and $P_S^6$ are now split quaternions.  The
cubic norm of $M(p)$ is given by
\begin{equation}
    \mathcal{N}(M(p)) = 3 \sqrt{3}\{ p^1 p^2 p^3 -
        \frac{1}{2} \left( p^1 P_S^4 \bar{P}_S^4 + p^2 P_S^5 \bar{P}_S^5 + p^3 P_S^6 \bar{P}_S^6 \right)
        + \frac{1}{\sqrt{2}} Re (P_S^4 P_S^5 P_S^6) \}
\end{equation}
where $Re(X_S)$ denotes the real part of $X _S\in \mathbb{H}_S$ and
$\bar{X}_S$ is the quaternion conjugate of $X_S$.  If we expand the
elements $P_S^4$, $P_S^5$ and $P_S^6$ in terms of their components
\begin{eqnarray}
 P_S^4 = p^4 + p^{7} j_1 +  p^{4+3m} j_m^s  \nonumber \\
 \bar{P}_S^4 = p^4 - p^{7} j_1 -  p^{4+3m} j_m^s \nonumber \\
 P_S^5 = p^5 + p^{8} j_1 +p^{5+3m} j_m^s\nonumber \\
 \bar{P}_S^5 = p^5 -p^{8} j_1 -p^{5+3m} j_m^s \\
 P_S^6 = p^6 + p^{9} j_1 + p^{6+3m} j_m^s\nonumber \\
 \bar{P}_S^6 = p^6 - p^{9} j_1 - p^{6+3m} j_m^s\nonumber
\end{eqnarray}
where the indices  $m,n,..$  are summed over and take values $2,3$. 

The cubic norm $\mathcal{N}(M(p))$ of $M(p)$ can thus be written as 
\begin{eqnarray} \label{cubicform_E7}
      \mathcal{N}(M(p))& = &3 \sqrt{3} \Bigg\{ p^1 p^2 p^3 -  \frac{1}{2} p^1 [( p^4)^2  + (p^{7})^2  - p^{4+3m} p^{4+3m}] \nonumber \\ &&
       -\frac{1}{2} p^2 [ (p^5)^2 + (p^{8})^2 - p^{5+3m} p^{5+3m} ] -\frac{1}{2} p^3 [ (p^6)^2 + (p^{9})^2 - p^{6+3m} p^{6+3m} ] \nonumber  \\ &&
      + \frac{1}{\sqrt{2}} \Big\{ p^4 p^5 p^6   - p^4 \left( p^{8} p^{9}- p^{(5+3m)} p^{(6+3m)} \right) - p^5 \left( p^{7}  p^{9} - p^{(4+3m)} p^{(6+3m)} \right)  \nonumber  \\ &&   - p^6 \left(p^{7}  p^{8} -p^{(4+3m)}  p^{(5+3m)}\right)\Big\} 
       +\frac{1}{\sqrt{2}} \epsilon_{1mn} p^{7} p^{5+3m} p^{6+3n} \\ \nonumber &&
        +\frac{1}{\sqrt{2}} \epsilon_{m1n} p^{4+3m} p^{8} p^{6+3n} 
         +\frac{1}{\sqrt{2}} \epsilon_{mn1} p^{4+3m} p^{5+3n} p^{9}       
      \Bigg\} \nonumber
\end{eqnarray}
With the above labelling of the basis elements we have 
\begin{eqnarray}
\eta_{AB}& =& \delta_{AB}  \\ \nonumber
\eta_{RS}& =& - \delta_{RS} \\
\eta_{AR}& =& 0 \nonumber
\end{eqnarray}
where $A,B,..= 1,2, ...,9$ and $ R,S,..= 10,11,...,15$.  The conjugate
elements $\Tilde{M}(p)$ of $J_3^{\mathbb{C}_S}$ obtained by
conjugation $*$ is given by
\begin{equation}
M (p)  = \frac{1}{\sqrt{2}}\left( {\begin{array}{*{20}c}
   \sqrt{2} p^1 &( P_S^6)^* & (\bar{P}_S^5)^*  \\
   (\bar{P}_S^6)^* & \sqrt{2}p^2 & (P^4)^*  \\
  ( P_S^5)^* & (\bar{P}_S^4)^* &\sqrt{2}p^3  \\
\end{array}} \right)
\end{equation}
where under the conjugation $*$ we have
\begin{eqnarray}
j_1^*=j_1 \\ \nonumber
(j_m^s)^* = -j_m^s 
\end{eqnarray}

The maximal compact subgroup of $E_{7(7)}$ is $SU(8)$ and its
generators can be obtained by restricting the range of indices in
\ref{maximalcompact_E8} such that $A,B,..= 1,2,...,9$ and $R,S,..=
10,11,...,15$. The quasiconformal compactification of the 33
dimensional space on which $E_{7(7)}$ acts is
\begin{equation}
  \mathcal{K}( QConf(J_3^{\mathbb{H}_S}) =\frac{SU(8)}{SU(4)\times SU(4)}
\end{equation}
The subgroup $SO(6,6)$ has three different embeddings in $E_{7(7)}$,
two of which are conformal acting on 15 coordinates or momenta,
respectively. The conformal compactification of the 15 dimensional
space on which $SO(6,6)$ acts is
\begin{equation}
\mathcal{K}(Conf(J_3^{\mathbb{H}_S}) = \frac{SU(4)\times SU(4)}{SU(4)} 
\end{equation}

We recall that for the Euclidean Jordan algebra $J_3^{\mathbb{H}}$ the associated quasiconformal group is $E_{7(-5)}$ and the 
quasiconformal  compactification of the corresponding 33 dimensional space is \cite{mg2009} 
%$J_3^{\mathbb{H}}$ are \eqn
\begin{equation}
\mathcal{K}(QConf( J_3^{\mathbb{H}}) = \frac{SO(12)\times SU(2)}{SU(6)\times U(1)}
\end{equation}
The quasiconformal group $E_{7(-5)}$ has a subgroup $SO^*(12)$ which
acts on the underlying Jordan algebra $J_3^{\mathbb{H}}$ as a
conformal group. The conformal compactification of the corresponding
15 dimensional space is
\begin{equation}
\mathcal{K}(Conf(J_3^{\mathbb{H}}) = \frac{SU(6)}{USp(6)} \times S^1 
\end{equation}

\subsection{Quasiconformal Group $E_{6(6)}$ }

Quasiconformal group associated with the Jordan algebra
$J_3^{\mathbb{C}_S} $ over the split complex numbers $\mathbb{C}_S$ is
$E_{6(6)}$. A general element of $J_3^{\mathbb{C}_S} $ can be written
in the form
\begin{equation}
M (p)  = \frac{1}{\sqrt{2}}\left( {\begin{array}{*{20}c}
   \sqrt{2} p^1 & P_S^6 & \bar{P}_S^5  \\
   \bar{P}_S^6 & \sqrt{2}p^2 & P^4  \\
   P_S^5 & \bar{P}_S^4 &\sqrt{2}p^3  \\
\end{array}} \right)
\end{equation}
where $P_S^4$, $P_S^5$ and $P_S^6$ are now split complex numbers.  The
cubic norm of $M(p)$ is given by
\begin{equation}
    \mathcal{N}(M(p)) = 3 \sqrt{3}\{ p^1 p^2 p^3 -
        \frac{1}{2} \left( p^1 P_S^4 \bar{P}_S^4 + p^2 P_S^5 \bar{P}_S^5 + p^3 P_S^6 \bar{P}_S^6 \right)
        + \frac{1}{\sqrt{2}} Re (P_S^4 P_S^5 P_S^6) \}
\end{equation}
where $Re(X_S)$ denotes the real part of $X _S\in \mathbb{C}_S$ and $\bar{X}_S$ is the complex  conjugate of $X_S$. 
If we expand the elements $P_S^4, P_S^5 $ and $P_S^6$ in terms of their  components
\begin{eqnarray}
 P_S^4 = p^4 + p^{7} j^s \nonumber \\
 \bar{P}_S^4 = p^4 - p^{7} j^s \nonumber \\
 P_S^5 = p^5 + p^{8} j^s\nonumber \\
 \bar{P}_S^5 = p^5 -p^{8} j^s \\
 P_S^6 = p^6 + p^{9} j^s\nonumber \\
 \bar{P}_S^6 = p^6 - p^{9} j^s\nonumber
\end{eqnarray}
where $j^s$  is the split ``imaginary'' unit that squares to +1
\begin{equation}
(j^s)^2= 1
\end{equation} 

The cubic norm $\mathcal{N}(M(p))$ of $M(p)$ can thus be written as
\begin{eqnarray} \label{cubicform_E6}
      \mathcal{N}(M(p))& = &3 \sqrt{3} \Bigg\{ p^1 p^2 p^3 -  \frac{1}{2} p^1 [( p^4)^2  - (p^{7})^2 ] \nonumber \\ &&
       -\frac{1}{2} p^2 [ (p^5)^2 - (p^{8})^2 ] -\frac{1}{2} p^3 [ (p^6)^2 - (p^{9})^2] \nonumber  \\ &&
      + \frac{1}{\sqrt{2}} \Big\{ p^4 p^5 p^6   + p^4 \left( p^{8} p^{9} \right) + p^5 \left( p^{7}  p^{9} \right)  \nonumber  \\ &&   + p^6 \left(p^{7}  p^{8} \right)\Big\}      
      \Bigg\} \nonumber
\end{eqnarray}
With the above labelling of the basis elements we have 
\begin{eqnarray}
\eta_{AB}& =& \delta_{AB}  \\ \nonumber
\eta_{RS}& =& - \delta_{RS} \\
\eta_{AR}& =& 0 \nonumber
\end{eqnarray}
where $A,B,..= 1,2, ...,6$ and $ R,S,..= 7,8,9$.  The conjugate
elements $\Tilde{M}(p)$ of $J_3^{\mathbb{H}_S}$ obtained by
conjugation $*$ is given by
\begin{equation}
M (p)  = \frac{1}{\sqrt{2}}\left( {\begin{array}{*{20}c}
   \sqrt{2} p^1 &( P_S^6)^* & (\bar{P}_S^5)^*  \\
   (\bar{P}_S^6)^* & \sqrt{2}p^2 & (P^4)^*  \\
  ( P_S^5)^* & (\bar{P}_S^4)^* &\sqrt{2}p^3  \\
\end{array}} \right)
\end{equation}
where under the conjugation $*$ we have
\begin{equation}
 (j^s)^*= - (j^s) 
\end{equation}

The maximal compact subgroup of $E_{6(6)}$ is $USp(8)$ and its
generators can be obtained by restricting the range of indices in
\ref{maximalcompact_E8} such that $A,B,..= 1,2, ...,6$ and $
R,S,..=7,8,9$. The quasiconformal compactification of the 21
dimensional space on which $E_{6(6)}$ acts is
\begin{equation}
\mathcal{K}( QConf(J_3^{\mathbb{C}_S}) =\frac{USp(8)}{SU(4)}
\end{equation}
The subgroup $SL(6,\mathbb{R})$ has three different embeddings in
$E_{6(6)}$, two of which are conformal acting on 9 coordinates or
momenta, respectively. The conformal compactification of the
corresponding 9 dimensional space on which $SL(6,\mathbb{R})$ acts is
\begin{equation}
\mathcal{K}(Conf(J_3^{\mathbb{C}_S}) = \frac{SO(6)}{SO(3)\times SO(3)} 
\end{equation}

We recall that for the Euclidean Jordan algebra $J_3^{\mathbb{C}}$ the associated quasiconformal group is $E_{6(2)}$ and the 
quasiconformal  compactification of the corresponding 21 dimensional space is \cite{mg2009} 
%$J_3^{\mathbb{H}}$ are \eqn
\begin{equation}
\mathcal{K}(QConf( J_3^{\mathbb{C}}) = \frac{SU(6)\times SU(2)}{SU(3)\times SU(3) \times U(1)}
\end{equation}
The quasiconformal group $E_{6(2)}$ has a subgroup $SU(3,3)$ which
acts on the underlying Jordan algebra $J_3^{\mathbb{C}}$ as a
conformal group. The conformal compactification of the corresponding 9
dimensional space is
\begin{equation}
\mathcal{K}(Conf(J_3^{\mathbb{C}}) = \frac{SU(3)\times SU(3)}{SU(3)} \times S^1 
\end{equation}

\subsection{ Quasiconformal Group $SO(n+3,m+3)$ }

For the generic nonsimple Jordan algebras $(\mathbb{R} \oplus
\Gamma_{(n,m)})$ of degree three the cubic form is
\begin{equation}
\mathcal{N}(q) =C^{IJK} q_Iq_Jq_K=  \frac{3\sqrt{3}}{2} q_1 [ 2q_2q_3 - \eta^{\hat{I}\hat{J}} q_{\hat{I}} q_{\hat{J}} ]
\end{equation}
where $\hat{I},\hat{J},...=4,5,..., (m+n+1) $ and  
\begin{eqnarray}
\eta^{\hat{I}\hat{J}} &= &\delta^{\hat{I}\hat{J}} \, , \,\,\,\, \hat{I},\hat{J} = 4,5,...,m+2 \\ \nonumber
\eta^{\hat{I}\hat{J}}& =& -\delta^{\hat{I}\hat{J}} \, , \,\,\,\, \hat{I},\hat{J} = m+3,m+4,...,n+m+1 
\end{eqnarray}
and the associated quasiconformal group is $SO(n+3,m+3)$.  The quartic
correction term $C^S_4$ that appears in $H^S_4$ in the spherical
vectors is given by
\begin{eqnarray}
        C^S_4 ( \mathbb{R}\oplus \Gamma_{(n,m)})  &= &
      - \frac{81}{2} \left\{ (p^2 - p^3)(q_2 - q_3) + 2 p^{\hat{I}} q_{\hat{I}} \right\}^2  \\ &&
    + \frac{81}{2} \left\{ (p^2 -p^3)^2 + 2 \eta_{\hat{I}\hat{J}} p^{\hat{I}} p^{\hat{J}} \right\}
     \left\{ (q_2 - q_3)^2  + 2 \eta^{\hat{I}\hat{J}} q_{\hat{I}} q_{\hat{J}} \right\} \nonumber
\end{eqnarray}
and the full metric $\eta_{IJ}$ is given by
\begin{eqnarray}
\eta_{AB} &= &\delta_{AB} \,\, , \,\, for \,\,  A,B,..= 1,2,...., m+2 \\ \nonumber
\eta_{RS} &= &-\delta_{RS} \,\, , \,\, for \,\, R,S,..= m+3, m+4,..., m+n+1 
\end{eqnarray}

The quasiconformal compactification of the $(2n+2m+5)$ dimensional
space on which $SO(n+3,m+3)$ is realized as the quasiconformal group
is
\begin{equation}
\mathcal{K}( QConf(\mathbb{R} \oplus \Gamma_{(n,m)}) = \frac{SO(n+3) \times SO(m+3)}{SO(n+1) \times SO(m+1) \times SO(2)}
\end{equation}
The subgroup $SO(m+1,n+1) \times SO(2,1) $ can be embedded in
$SO(n+3,m+3)$ in three inequivalent ways. Two of these embeddings act
nonlinearly as conformal groups on the coordinates and momenta ,
respectively. The conformal compactification of the $(m+n+1)$
dimensional space on which $SO(m+1,n+1) \times SO(2,1) $ acts via
conformal transformations is
\begin{equation}
\mathcal{K}( Conf( \mathbb{R} \oplus \Gamma_{(n,m)})) =\frac{SO(m+1)\times SO(n+1)}{SO(m) \times SO(n)} \times S^1 
\end{equation}

For Euclidean Jordan algebras $(\mathbb{R} \oplus \Gamma_{(1,m)}$ the
quasiconformal group is $SO(4,m+3)$ and the conformal group is
$SO(2,m+1)\times SO(2,1)$ leading to the following compactified spaces
\begin{equation}
\mathcal{K}( QConf(\mathbb{R} \oplus \Gamma_{(1,m)}) = \frac{SO(4) \times SO(m+3)}{SO(2) \times SO(m+1) \times SO(2)}
\end{equation}
\begin{equation}
\mathcal{K}( Conf(\mathbb{R} \oplus \Gamma_{(1,m)})) =\frac{SO(m+1)}{SO(m)} \times S^1 \times S^1 
\end{equation}

\section{ Exceptional $N=2$ MESGT , $N=8$ Supergravity and Jordan Algebras of Degree Four}
As explained above the U-duality symmetry groups of the maximal
supergravity in $5,4$ and $ 3$ dimensions are simply the Lorentz 
(reduced structure), conformal and quasiconformal groups of the split
exceptional Jordan algebra $J_3^{\mathbb{O}_S}$, which is not
Euclidean. On the other hand the corresponding U-duality groups of the
exceptional $N=2$ supergravity are the Lorentz, conformal and
quasiconformal groups of the real exceptional Jordan algebra
$J_3^{\mathbb{O}}$ which is Euclidean.  The scalar manifold of the
exceptional theory in five dimensions is simply
\begin{equation*}
 \mathcal{M}_5(J_3^{\mathbb{O}}) = \frac{Str_0(J_3^{\mathbb{O}})}{Aut(J_3^{\mathbb{O}})}= \frac{E_{6(-26)}}{F_4} 
\end{equation*}
However the scalar manifold of the maximal supergravity is 
\begin{equation*} 
  \frac{E_{6(6)}}{USp(8)}
\end{equation*}
so that the maximal compact subgroup is not the automorphism group of
the split exceptional Jordan algebra. Remarkably, there is a
formulation of the exceptional Jordan algebras in terms of degree four
Jordan algebras in which the situation gets reversed. As was shown in
\cite{Gunaydin:2005df} the exceptional supergravity can also be
formulated in terms of Lorentzian Jordan algebra
$J_{(1,3)}^{\mathbb{H}}$ of $4\times 4$ quaternionic matrices that are
Hermitian with respect to a Lorentzian metric. This is achieved by the
mapping between the traceless elements , which we denote as
$J_{(1,3)_0}^{\mathbb{H}}$, of $J_{(1,3)}^{\mathbb{H}}$ and the
elements of the exceptional Jordan algebra $J_3^{\mathbb{O}}$:
\begin{equation}
\mathbf{x}  \in J_3^{\mathbb{O}}  \Longrightarrow \mathbf{X} \in J_{(1,3)_0}^{\mathbb{H}}
\end{equation}
such that the cubic norm $\mathcal{N}(\mathbf{x})$ of $\mathbf{x}$ is equal to the trace of $ \mathbf{X}^3 $:
\begin{equation}
\mathcal{N}(\mathbf{x}) = Tr ( \mathbf{X}^3) 
\end{equation}
Now the automorphism group of $ J_{(1,3)_0}^{\mathbb{H}} $ is
$USp(6,2)$ which is the manifest symmetry of the trace form. Hence the
trace of $\mathbf{X}^3$ has "hidden" extra symmetries which extend
$USp(6,2)$ to $E_{6(-26)}$, which is the Lorentz (reduced structure)
group of $J_3^{\mathbb{O}}$. However, the Lorentz (reduced structure)
group $SU^*(8)$ of $J_{(1,3)}^{\mathbb{H}}$ is not a symmetry of the
exceptional MESGT in five dimensions.

Similarly there is a mapping between the elements $\mathbf{y}$ of the
split exceptional Jordan algebra $J_3^{\mathbb{O}_S}$ and the
traceless elements $\mathbf{Y}$ of Euclidean Jordan algebra
$J_4^{\mathbb{H}}$ of $4\times 4$ Hermitian matrices over the division
algebra of quaternions $\mathbb{H}$ \cite{MR735416}
\begin{equation}
\mathbf{y}  \in J_3^{\mathbb{O}_S}  \Longrightarrow \mathbf{Y} \in J_{4_0}^{\mathbb{H}}
\end{equation}
such that one finds
\begin{equation}
\mathcal{N}(\mathbf{y}) = Tr ( \mathbf{Y}^3) 
\end{equation}
The automorphism group $USp(8)$ of $J_4^{\mathbb{H}}$ is the manifest
symmetry of the trace form. Again extra hidden symmetries of
$Tr(\mathbf{Y}^3)$ extend it to the Lorentz (reduced structure) group
$E_{6(6)}$ of $J_3^{\mathbb{O}_S}$. However the scalar manifold of the
maximal supergravity in five dimensions is
\begin{equation}
\mathcal{M}_5(N=8) = \frac{Str_0(J_3^{\mathbb{O}_S})}{Aut(J_{4}^{\mathbb{H}})}=\frac{E_{6(6)}}{USp(8)}
\end{equation}
The Lorentz (reduced) structure group $SU^*(8)$ of $J_4^{\mathbb{H}}$
is not a symmetry of maximal supergravity in five dimensions.  In four
dimensions the scalar manifold of maximal supergravity is
\begin{equation}
\mathcal{M}_4(N=8) = \frac{Conf(J_3^{\mathbb{O}_S)}}{\widetilde{Str}_0(J_4^{\mathbb{H}})} = \frac{E_{7(7)}}{SU(8)} 
\end{equation}
where $\widetilde{Str}_0(J_4^{\mathbb{H}})$ denotes the compact real
form of the Lorentz (reduced structure) group of
$J_4^{\mathbb{H}})$. In three dimensions one has the scalar manifold
\begin{equation}
\mathcal{M}_3 (N=8) = \frac{QConf(J_3^{\mathbb{O}_S})}{\widetilde{Conf}(J_4^{\mathbb{H}})}=\frac{E_{8(8)}}{SO(16)}
\end{equation}
where $\widetilde{Conf}(J)$ refers to the compact real form of the conformal group of $J$.

One can truncate the above correspondences between the real and split exceptional Jordan algebras and $J_{(1,3)}^{\mathbb{H}}$ and $J_4^{\mathbb{H}}$ , respectively, to correspondences between rank three complex and quaternionic Jordan algebras and rank four real and complex Jordan algebras:
\eqn
J_3^{\mathbb{C}} & \Longleftrightarrow J_{(1,3)_0}^{\mathbb{R}} \nonumber \\
J_3^{\mathbb{H}} &\Longleftrightarrow J_{(1,3)_0}^{\mathbb{C}} \nonumber \\
J_3^{\mathbb{C}_S}& \Longleftrightarrow  J_{4_0}^{\mathbb{R}} \nonumber \\
J_3^{\mathbb{H}_S} &\Longleftrightarrow  J_{4_0}^{\mathbb{C}}
 \enn

Remarkably the $N=2$ MESGT theories defined by Lorentzian Jordan algebras $J_{(1,3)}^{\mathbb{A}}$ belong to three infinite families of unified MESGT's in five dimensions defined by Lorentzian Jordan algebras of arbitrary rank \cite{Gunaydin:2005df}. Study of quasiconformal groups associated with Lorentzian Jordan algebras will be left to future studies.

%%%%%%%%%%%%%%%%%%%%%%%%%%%
%%%%%%%%%%%%%%%%%%%%%%%%%
{\bf Acknowledgement:}
 This work was supported in part by the National Science
Foundation under grant number PHY-0555605. Any opinions, findings and
conclusions or recommendations expressed in this material are those of
the authors and do not necessarily reflect the views of the National
Science Foundation.

%.
%\bibliography{QCGTWISTOR_HEPTH2}
%\bibliographystyle{utphys}
\bibliography{QCGTWISTOR_SPLIT_FINAL}
\bibliographystyle{utphys}
\end{document}